\documentclass[aps,prb,twocolumn,showpacs,superscriptaddress,groupedaddress]{revtex4-1}
\usepackage{graphicx}
\usepackage{amsmath}
\usepackage{amssymb,color}
\usepackage[utf8]{inputenc}
\usepackage{braket}
\input{epsf}

\usepackage{bbm}
\usepackage{array}
\usepackage[markup=nocolor]{changes}
\usepackage{braket}
\usepackage{float}
\input{epsf}
\newcommand{\beq} {\begin{equation}}
\newcommand{\eeq} {\end{equation}}
\newcommand{\bea} {\begin{eqnarray}}
\newcommand{\eea} {\end{eqnarray}}
\newcommand{\be} {\begin{equation}}
\newcommand{\ee} {\end{equation}}

\DeclareMathOperator{\arctanh}{tanh^{-1}}

\definecolor{darkgreen}{RGB}{0,170,0}

\newcommand{\nn}{\nonumber}
\newcommand{\kk}{\textbf{k}}
\newcommand{\Q}{\textbf{Q}}
\newcommand{\pp}{\textbf{p}}

\begin{document}
\title {Specific Heat and the gap structure of a Nematic Superconductor, application to FeSe}
\author{Kazi Ranjibul Islam$^{1*}$ }
\author{Jakob B\"oker$^{2*}$}
\author{Ilya M. Eremin$^2$}
\author{Andrey V. Chubukov$^1$}
\affiliation{1-School of Physics and Astronomy and William I. Fine Theoretical Physics Institute,
University of Minnesota, Minneapolis, MN 55455, USA}
\affiliation{2-Institut f\"ur Theoretische Physik III, Ruhr-Universit\"at Bochum, 44801 Bochum, Germany}

\date{\today}
\begin{abstract}
We report the results of our in-depth analysis of spectroscopic  and  thermodynamic  properties of a multi-orbital metal, like FeSe, which first develops a nematic order and then undergoes a transition into a superconducting state, which co-exists with nematicity.
We analyze the angular dependence of the gap function and specific heat $C_V (T)$ of such nematic superconductor. We specifically address three issues: (i) angular dependence of the gap in light of the competition between nematicity-induced $s$-$d$ mixture and orbital transmutation of low-energy excitations in the nematic state, (ii) the effect of nematicity on the magnitude of the jump of the specific heat $C_V (T)$ at $T_c$ and the temperature dependence of $C_V (T)$ below $T_c$, and (iii) a potential transition at $T_{c1} < T_c$ from an $s+d$ state to an $s + e^{i\eta} d$ state that breaks time-reversal symmetry. We consider two scenarios for a nematic order: scenario A, in which this order develops between $d_{xz}$ and $d_{yz}$ orbitals on hole and electron pockets  and scenario B, in which there is an additional component of the nematic order for $d_{xy}$ fermions on the two electron pockets.
   \end{abstract}
\maketitle

\section{ Introduction.}
Iron-based unconventional superconductors demonstrate remarkable properties, which include  multi-orbital low-energy electronic states and ubiquity of the nematic phase. A particularly interesting situation occurs when superconductivity is preceded by the development of a
nematic order  that breaks C$_4$ lattice rotational symmetry down to C$_2$. The most prominent example of this so-called nematic superconductor is FeSe, in which a nematic order develops at  $T_n \sim 90K$ at ambient pressure, while superconductivity develops at a much lower $T_c \sim 9K$, out of a nematic state~[\onlinecite{nematic_FeSe},\onlinecite{nem_review_5}].  Nematic superconductivity has been observed also in other Fe-based materials, but there the difference between $T_n$ and $T_c$ is much smaller~[\onlinecite{fernandes2016low}]. It has  been also argued that in some cases a nematic order does not exist in the normal state but is induced by superconductivity. A candidate for such behavior in the Fe-family is LiFeAs~[\onlinecite{kushnirenko2020nematic}]; the same behavior has been reported in twisted bilayer graphene~[\onlinecite{graphene_nematic}] and in doped topological insulator R$_x$Bi$_2$Se$_3$ (R=Cu, Nb, and Sr)~[\onlinecite{Matano2016,Yonesawa2017,Pan2016,Asaba2017}]. In this work, we focus on the theoretical analysis of the  spectroscopic and the thermodynamic  properties of such a nematic superconductor using  the case of FeSe, where $T_n$ is substantially larger than $T_c$.\newline
\begin{figure}[t]
		\centering
		\includegraphics[width=1\linewidth]{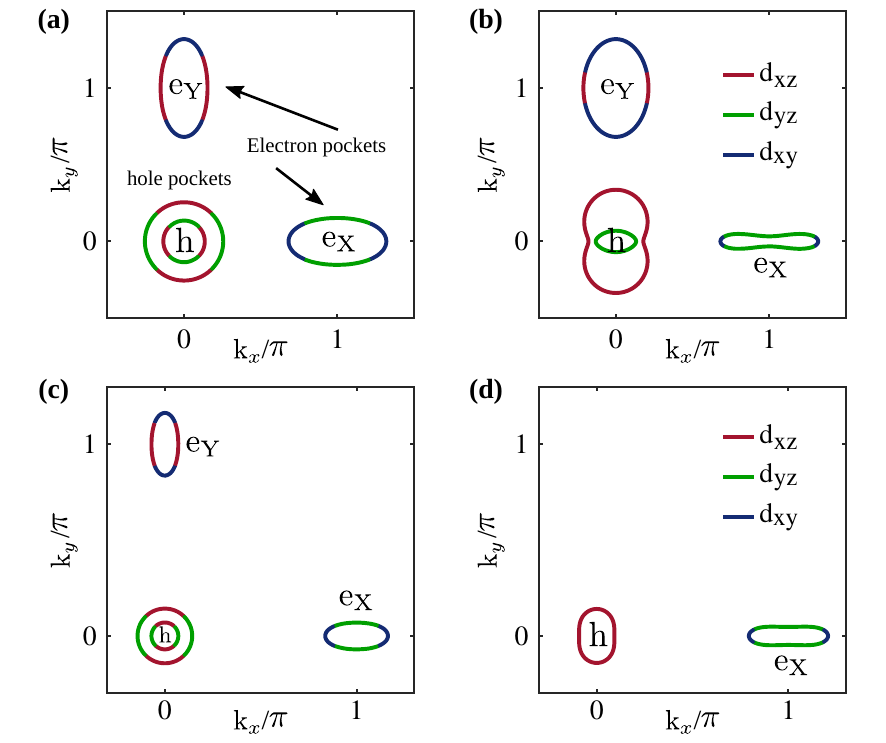}
		\caption{Fermi surface topology in 1-Fe unit cell of FeSe in  the tetragonal phase (a,c) and  the  orthorhombic (nematic) phase (b,d). The Fermi surface  evolution in (b) results from sign-changing nematic order involving $d_{xz}$ and $d_{yz}$ orbitals, in (d) it additionally involves sizable non-local $d_{xy}$ nematic order. We refer to panels (b) and (d) as "scenario A" and "scenario B", respectively. The color code follows the major orbital content. Fitting parameters for (a,b) are taken from Refs.~[\onlinecite{fernandes2016low},\onlinecite{fernandes2014distinguishing}] and for (c,d) from  Ref.~[\onlinecite{rhodes2020non}].}
		\label{Fig:0}
\end{figure}
The electronic structure of FeSe in the tetragonal phase consists of two hole pockets, centered around the $\Gamma$ point (the inner one and the outer one), and electron pockets, centered around the $X$ and the $Y$ points of the Brillouine Zone, respectively (Fig.~\ref{Fig:0}(a)). Here, we use the notation of the 1-Fe unit cell. The hole pockets and the corresponding bands are composed of fermions from $d_{xz}$ and $d_{yz}$ orbitals, the $X$-pocket/band is a mixture of $d_{yz}$ and $d_{xy}$ orbitals, and the $Y$-pocket/band is a mixture of $d_{xz}$ and $d_{xy}$ orbitals.  ARPES studies revealed that in FeSe the inner hole pocket is quite small in the tetragonal phase and disappears in the presence of a nematic order, when the corresponding band sinks below the Fermi level~[\onlinecite{Watson2015}] (Fig.~\ref{Fig:0}(b)). The inner hole band then does not affect system behavior at low energies and we neglect it in our analysis. For the  outer hole pocket, the orbital content in the tetragonal phase is predominantly $d_{xz}$ along the $k_y$-direction and $d_{yz}$ along the $k_x$-direction.\newline

We consider two scenarios for the nematic order, $\Phi$. In the first (scenario A) we assume that $\Phi$  splits the occupations of $d_{xz}$ and $d_{yz}$ orbitals:
\beq
\Phi_{xz/yz} = \langle d^\dagger_{xz} d_{xz} - d^{\dagger}_{yz} d_{yz} \rangle
\label{a_1}
\eeq
Furthermore, we follow earlier theoretical and experimental studies~[\onlinecite{chubukov2016magnetism, onari2016sign, fanfarillo2016orbital, benfatto2018nematic,udina}], which showed that such $\Phi$ changes sign between hole and electron pockets.  We label $\Phi$ on the outer hole pocket as $\Phi_h$ and the one on the $Y$ and the $X$ electron pockets as $\Phi_e$(sgn $\Phi_e=-$sgn $\Phi_h $). In the second scenario (scenario B), we assume that in addition to $\Phi_{h,e}$,  nematicity gives rise to a substantial
difference between occupations of $d_{xy}$ fermions  on the $Y$ and  the $X$ pockets~[\onlinecite{chubukov2016magnetism,rhodes2020non,laura,*laura_2,*laura_3,*laura_4}]. The corresponding nematic order parameter is then given by
\beq
 \Phi_{xy}=\langle d^\dagger_{xy,Y} d_{xy,Y}-d^{\dagger}_{xy,X}d_{xy,X}\rangle.
 \label{a_1_1}
\eeq
Scenario B  was recently advanced in Ref. [\onlinecite{rhodes2020non}]  as a way to explain the thermal evolution of the band structure across the tetragonal to orthorhombic transition as well as the fact that ARPES and QPI measurements in the nematic phase detect a peanut-shaped $X$  pocket, but  do not see the Y pocket~[\onlinecite{Watson2017b,Yi2019,Huh2020,Cai2020,Cai2020b,Rhodes2020}].
The argument here is that for large enough $\Phi_{xy}$, the $Y$ pocket disappears, as its bottom moves above the Fermi level (Fig.~\ref{Fig:0}(d)). A similar behavior has been obtained in  monoclinic systems by allowing a non-zero interorbital d$_{xz}$--d$_{xy}$ and d$_{yz}$--d$_{xy}$ nematicity~[\onlinecite{Steffensen2021}]. Within scenario A, it was argued~[\onlinecite{lanata2013orbital}] that the $Y$ pocket is not observed, because in the nematic phase it becomes predominantly $d_{xy}$ (the blue ellipse in Fig.~\ref{Fig:0}(b)), and these excitations are less coherent than the ones for $d_{xz}$ and $d_{yz}$ fermions~[\onlinecite{medici}]. In this work we analyze the effect of nematicity on the superconducting state within both scenarios. We discuss the angular dependence of the superconducting gap, most notably on the hole pockets, and the  behavior on the specific heat $C (T)$ at and below $T_c$.\newline

Multi-orbital superconductivity in Fe-based materials in the absence of a nematic order has been extensively studied by many groups~[\onlinecite{nem_review,nem_review_1,nem_review_2,nem_review_3,nem_review_4,nem_review_5}].
A mixed orbital content of low-energy excitations  implies  that the pairing interaction necessarily  has two orthogonal components:  $s$-wave and $d$-wave, even when the  interaction is local in the orbital basis. An $s$-wave interaction is attractive in the $s^{+-}$ sub-channel (the sign of the gap on the hole  pocket is opposite to that on electron $X$ and $Y$ pockets), a $d$-wave interaction is attractive in the $d_{x^2-y^2}$ sub-channel (the gap on the hole pocket scales as $\cos{2\theta}$, where
$\theta$ is the angle along the pocket, and has four nodes, while the gaps on the $X$ and the $Y$ pocket  have opposite sign). In both cases, the gaps on the $X$ and $Y$ pockets are sign-preserving, but generally have minima at the points where $d_{xz}$ ($d_{yz}$) orbital content vanishes. These minima can become nodes if $d_{xy}$ orbitals contribute to superconductivity~[\onlinecite{rhodes2020non}].\newline

The pairing interaction in the $s$-wave and the $d$-wave channels is expressed in terms of dressed interactions between hole and electron pockets:  intra-orbital density-density interaction $U_{he}$, and inter-orbital pair-hopping interactions  $J_{he}$ and $J_{ee}$ (see Sec \ref{sec:pairing} below).
The terms $U_{he}$ and $J_{he}$  are enhanced by magnetic fluctuations with momenta near $(0,\pi)$ and $(\pi,0)$ (the distances between the centers of the $\Gamma$ and the $X$ and the $Y$ pockets, respectively), and $J_{ee}$ is enhanced by magnetic fluctuations with momentum $(\pi,\pi)$ (the distance between the $X$ and the $Y$ pockets). We follow earlier works~[\onlinecite{kang2018superconductivity},\onlinecite{kang2018time}] and assume that the dressed pairing interaction in the tetragonal phase is somewhat stronger in the $s^{+-}$ channel. This implies that the pairing state without nematic order would be $s^{+-}$.\newline

Superconductivity in the presence of a small nematic order $\Phi_{h,e}$ has been studied previously in  Refs.~[\onlinecite{kang2018superconductivity,kang2018time,davis,*davis_1}].
The expected outcome is that a nematic order mixes $s$-wave and $d$-wave pairing channels, creating a mixed $s+d$ state. A general belief, coming from small $\Phi_{h,e}$ analysis is that in  such a state the gap  along the hole pocket is $\Delta_h (\theta) = \Delta_s + \Delta_d \cos{2\theta}$, where $\theta$ is the angle along the pocket. The magnitude of $\Delta_d$ increases with $\Phi$, and if one would extend the small $\Phi$ analysis to larger $\Phi$, one would obtain that $\Delta_h (\theta)$  develops a deep minima and then accidental nodes. This reasoning has been applied to explain  ARPES and STM data in FeSe~[\onlinecite{ARPES,*ARPES_1,davis}]. We argue that this is not necessarily the case  because there is a second, competing effect of nematicity. Namely, a nematic order changes  the orbital composition of the pockets (this phenomenon has been termed orbital transmutation~[\onlinecite{udina}]). This leads to two effects. First, the variable $\theta$ gets renormalized and becomes dependent on $\Phi_h$. At large enough $\Phi_h$, the dressed $\theta$ (called $\phi$ later in the paper) clusters near $\pm \pi/2$, depending on the sign of $\Phi_h$, and the gap looses its  angle dependence. Second, the ratio  $\Delta_d/\Delta_s$ becomes a non-linear function of $\Phi_h$, and the ratio $\Phi_h/\Phi_e$. Furthermore, in some intervals of  $\Phi_h/\Phi_e$ it remains below one even at large $\Phi_h$ values.  This prevents the appearance of the nodes even if the angular variation of the d-wave gap component is still a sizable one.\newline

Our goal is to understand what happens at intermediate values of $\Phi$, relevant to FeSe, in particular, whether there exists the range of $\Phi_h$ and $\Phi_h/\Phi_e$,  where $\Delta_h$ has nodes. We show that this range exists, but is confined to near-equal interactions in $s$-wave and $d$-wave channels. Nevertheless, even if the gap does not have nodes, its angular variation follows the orbital content of the hole pocket and undergoes a strong evolution once the orbital content changes. For completeness, we also consider the case when the $d$-wave interaction is stronger than the one in the $s^{+-}$ channel.  In this case, the gap has 4 nodes at small $\Phi_{h,e}$ and no nodes at large $\Phi_{h,e}$, due to orbital transmutation. We show that the transformation of the nodal structure at intermediate $\Phi_{h,e}$ is rather involved, and for some $\Phi_e/\Phi_h$ there exists an intermediate gap configuration with 8 nodes.\newline

We next consider the behavior of the specific heat $C_v(T)$ at and below $T_c$. We analyze how the jump of $C_v(T)$ varies with the type of nematic order and whether the jump primarily comes from fermions from $d_{xz}$ and $d_{yz}$ orbitals, or there is a sizable contribution from the $d_{xy}$ orbital. A similar issue has been recently studied~[\onlinecite{chichinadze2019specific}] for KFe$_2$As$_2$. There, $d_{xy}$ orbital gives the dominant contribution to $C_v(T)$ in the normal state because of large mass of $d_{xy}$ fermions, but contributes little to the jump of  $C_v(T)$ and also to temperature dependence of $C_v(T)$ in a wide temperature region below $T_c$, because a superconducting gap on this orbital is inversely proportional to its mass and is much smaller than the ones on $d_{xz}$ and $d_{yz}$ orbitals. We analyze whether the same holds for FeSe, using the values of quasiparticle masses,  extracted from ARPES. We find that the jump of the specific heat at $T_c$ is smaller than in BCS theory for the same number of pockets, by the same reason as in KFe$_2$As$_2$, $d_{xy}$ fermions substantially contribute to $C_v(T)$ in the normal state but little to the jump of $C_v(T)$ at $T_c$. We decompose $\delta C_v$ into contributions from different pockets and show that the largest contribution comes from fermions on a hole pocket in scenario A and from an electron pocket in scenario B. We analyze how $\delta C_v$  evolves with nematic order and again find strong correlation with the orbital transmutation.\newline

Finally, we address the issue of potential second transition to the new phase  within the superconducting state. The argument here is that in a situation, when the attraction in the $d_{x^2-y^2}$ channel is comparable to that in the $s^{+-}$ channel, a bi-quadratic coupling between $s$- and $d$-order parameters may turn the $s+d$ pairing state into an $s +i e^{i\eta} d$ state (the analog of a mixed $s + id$ state in the absence of nematicity). Such a state breaks $Z_2$ time-reversal symmetry, as the relative factor can be either $i$ or $-i$.  Recent specific heat measurements, $C_v(T)$, of FeSe~[\onlinecite{chen2017highly,sun2017gap,Sun2018,Jiao2017,Klein2019,Klein_Private_Commun}] found an anomaly at $T \sim 1K$, which might indicate the emergence of $s+ e^{i\eta} d$ order~[\onlinecite{kang2018time}]. To verify the scenario, we vary the relative strength of the pairing interactions in $s$-wave and $d$-wave channels  and analyze the Landau functional including both the bi-quadratic couplings between $s$- and $d$-gap components and the effect of orbital transmutation in the nematic phase. Although the orbital transmutation  shrinks the parameter range of $s+ e^{i\eta} d$ state, a transition into an  $s+ e^{i\eta} d$ state below $T_c$ is still possible.\newline

The structure of the paper is the following.  In the next Section we briefly discuss the electronic structure of FeSe.  In Sec.~\ref{pairinginteraction} we obtain the pairing interaction within scenario A, convert it into the band basis, and solve for the pairing gaps on hole and electron pockets.  In Sec.~\ref{gapequation} we analyze the angular dependence of the gap on the hole pocket  at various $\Phi_h$ and $\Phi_h/\Phi_e$. In  Sec.~\ref{sec:gap_T} we study temperature dependence of the gap below $T_c$. In Sec.~\ref{sec:spec_heat} we compute the jump of the specific heat at $T_c$ within both scenarios and compare them to the available experimental data. We decompose the jump into contributions from different orbitals and study their relative strength.  We also compute specific heat at $T < T_c$. In Sec.~\ref{sec:spec_heat_2} we  consider a putative transition into $s+ e^{i\eta} d$ state.  Finally, We present our conclusions in Sec.~\ref{sec:conclusion}.

\section{The band Hamiltoninan\label{sec:H}}
As mentioned in the Introduction, we consider a two-dimensional 3 band/3 pocket model Hamiltonian with a hole pocket, centered at the $\Gamma$ point of the BZ and two electron pockets, centered at $X=(0,\pi)$ and $Y=(\pi,0)$ points of the Brillouin zone, respectively. For simplicity, we neglect the effect of spin-orbit coupling on the band dispersion. The hole pocket and the corresponding hole band is composed of $d_{xz}$ and $d_{yz}$ orbitals. The $X-$ pocket/band is composed of $d_{yz}$ and $d_{xy}$ orbitals, and the $Y-$ pocket/band is composed of $d_{xz}$ and $d_{xy}$ orbitals. We introduce two-component spinors $\psi_{\Gamma}=\left(d_{xz},d_{yz}\right)^T$ and $\psi_{X/Y}=\left(d_{yz/xz},d_{xy}\right)^T$ and write the kinetic energy  $H_0$  as%
\begin{equation}
H_{0}=H_{\Gamma}+H_X+H_Y,
\label{noninteracting hamiltonian}
\end{equation}
where each term is bilinear in spinors. For scenario A we introduce the nematic order $\Phi$ as the difference in the occupation of $d_{xz}$ and $d_{yz}$ orbitals, see Eq.~(\ref{a_1}).  We define $\Phi$ on the hole pocket as $\Phi_h$ and on the electron pocket as $\Phi_e$. The latter is the difference in the occupation of $d_{xz}$ orbital on the $Y$ pocket and $d_{yz}$ orbital on the $X$ pocket. For scenario B we additionally introduce  a second component of a nematic order as the difference between occupations of $d_{xy}$ orbitals on $Y$ and $X$ pockets, see Eq.~(\ref{a_1_1}).
\subsection{Hole Pocket}
The band Hamiltonian for the hole pocket $H_{\Gamma}$ is~[\onlinecite{cvetkovic2013space,kang2018superconductivity,udina,morten}]
\begin{widetext}
\begin{equation}
 H_{\Gamma}=\psi_{\Gamma}^{\dagger} \left[\left(\mu_h-\frac{\kk^2}{2m_h}\right)\tau_0-\left(\frac{b}{2}\kk^2 \cos{2\theta_h}-\Phi_h\right)\tau_3 -c\kk^2\sin{2\theta_h}\tau_1\right]\psi_{\Gamma},
 \label{hole hamiltonian}
\end{equation}
\end{widetext}
where $\theta_h$ is the polar angle for momentum $\kk$, measured from the $k_x$-direction in the anti-clockwise direction. We set $c=-\frac{b}{2}$, which yields circular hole pockets in the tetragonal phase. The parameters of Eq.~(\ref{hole hamiltonian}) are listed in Table \ref{table:Table S1}, and  were obtained in Refs.~[\onlinecite{fernandes2016low},\onlinecite{fernandes2014distinguishing}] from fitting to ARPES data for FeSe at $k_z=\pi$.
\begin{table}[b]
\centering
\begin{tabular}{|c|c|c|}
\hline
  $\mu_h$ & $(2m_h)^{-1}$ & b \\
  \hline
   13.6  & 473  & 529 \\
   \hline
     \end{tabular}
     \caption{Band Parameters for the hole pocket}
     \label{table:Table S1}
\end{table}
Diagonalizing  Eq.~(\ref{hole hamiltonian}), we obtain two dispersions. In the absence of nematicity, they give rise  to the outer and the inner hole pockets, Fig.~\ref{Fig:1_1A}(a). At a finite $\Phi_h>\mu_h$, the inner hole pocket becomes very shallow and then disappears as the corresponding dispersion sinks below the Fermi level. For this reason, we  neglect the inner hole band in our analysis of the low-energy physics.

\begin{figure}[t]
\includegraphics[width=\columnwidth]{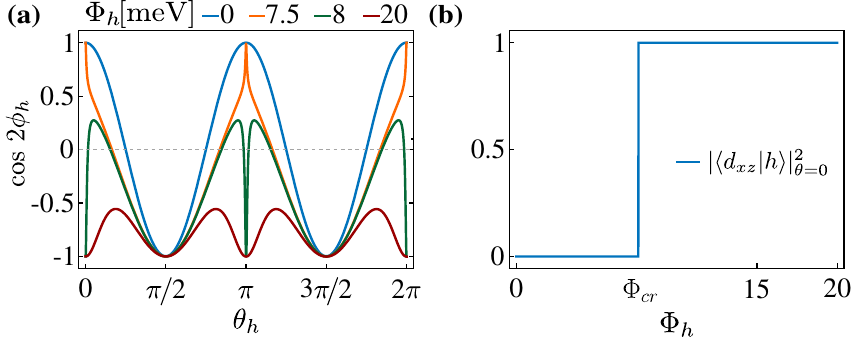}
\caption{(a) Angular variation of $\cos{2\phi_h}$ along  the hole Fermi pocket for selected values of $\Phi_h$, (b)  the $d_{xz}$ orbital weight at $\theta_h=0$ as a function of $\Phi_h$.}
\label{gapalongx}
\end{figure}

 The larger hole Fermi surface pocket survives at a finite $\Phi_h$ and becomes elliptical. The dispersion of the corresponding band is
 \begin{equation}
 \xi_h(\kk)=\mu_h-\frac{\kk^2}{2m_h}+\sqrt{\Phi_h^2+b^2 \frac{\kk^4}{4}-b \kk^2 \Phi_h \cos{2\theta_h}},
 \label{hole disperison}
\end{equation}
see Fig.~\ref{Fig:1_1A}(b). The band operator $h$ is a linear combination of fermionic operators from $d_{xz}$ and $d_{yz}$ operators:
\begin{equation}
h=\cos{\phi_h} d_{yz}+\sin{\phi_h} d_{xz},
\label{hole band operator}
\end{equation}
 where the momentum label ($\kk$) is implicit and $\phi_h$ is defined via
\begin{equation}
\cos{2 \phi_h}=\frac{b \frac{\kk^2}{2}\cos{2\theta_h}-\Phi_h}{\sqrt{\Phi_h^2+b^2 \frac{\kk^4}{4}-b \kk^2 \Phi_h \cos{2\theta_h}}}.
\label{cos2phih}
\end{equation}
 At $\Phi_h =0$, $\phi_h = \theta_h$, and $d_{yz}$ and $d_{xz}$ fermions contribute to $h$ with weights $|\langle d_{yz}|h\rangle|^2=\cos^2{\theta_h}$ and $|\langle d_{xz}|h\rangle|^2=\sin^2{\theta_h}$, simply related by $\pi/2$ rotation. At a non-zero $\Phi_h$, $\phi_h$ becomes different from $\theta_h$, and the weight of the two orbitals is no longer equal.  At large $\Phi_h$, $\cos{2 \phi_h} =- \text{sgn} \Phi_h$.  Choosing for definiteness $\Phi_h >0$, we find that $\phi_h = \pi/2$, hence the band operator $h$ in Eq.~(\ref{hole band operator}) becomes entirely $d_{xz}$, i.e., the hole pocket becomes mono-orbital.  This effect has been dubbed orbital transmutation~[\onlinecite{udina,morten}]. The angular variation of $\cos{2 \phi_h}$ on the outer hole pocket for intermediate values of $\Phi_h$ is shown in Fig.~\ref{gapalongx}(a). At $\Phi_h=\Phi_{cr}= bk^2_F/2 = \mu_h m_h b$, $\cos{2\phi_h}$ along $k_x$-direction jumps discontinuously from $+1$ to $-1$ (yellow and green curves in Fig.~\ref{gapalongx}(a)), and the orbital content jumps from a pure $d_{yz}$ to a pure $d_{xz}$ (see Fig.~\ref{gapalongx}(b)). Because of that jump, the angular average of $\cos{2\phi_h}$ and $\cos^2{2\phi_h}$ along the hole Fermi surface, viewed as a function of $\Phi_h$,  becomes non-analytic at $\Phi_{cr}$. In addition,  at $\Phi \geq \Phi_{cr}$, the number of nodes  of $\cos{2\phi_h}$ on the Fermi surface increases from $4$ to $8$ (the green curve in Fig.~\ref{gapalongx}(a)). We will show later that both features affect the structure of the superconducting gap function.  For  band parameters from Table \ref{table:Table S1}, $\Phi_{cr}\approx 7.6$ meV.

\subsection{X and Y Pockets}

The electron pockets are described by the band Hamiltonian $H_{X/Y}$~[\onlinecite{cvetkovic2013space,kang2018superconductivity,udina,morten}].
\begin{equation}
H_{X/Y}=\psi_{X/Y}^{\dagger}\begin{pmatrix}
A_{X/Y}^{(1)} & -i V_{X/Y}\\
i V_{X/Y} & A_{X/Y}^{(2)}
\end{pmatrix} \psi_{X/Y}.
\label{electro hamiltonian}
\end{equation}
The diagonal elements are
 \begin{align}
A_{X/Y}^{(1)}&=\frac{\kk^2}{2m_1}-\mu_1-\frac{a_1}{2}\kk^2 \cos{2\theta_{X/Y}}\pm \Phi_e,\\
A_{X/Y}^{(2)}&= \frac{\kk^2}{2m_3}-\mu_3-\frac{a_3}{2}\kk^2 \cos{2\theta_{X/Y}}.
\end{align}
Here, $\kk$ is measured from $X=\left(\pi,0\right)$ for the X pocket, and from  $Y=\left(0,\pi\right)$ for the $Y$ pocket and the upper (lower) sign corresponds to the X(Y)-pocket. $\theta_X(\theta_Y)$ is the polar angle, measured with respect to $k_{x(y)}$-direction for the $X(Y)$ electron pocket in the anti-clockwise direction. $\Phi_e$ is the electron nematic order defined as,
  $\Phi_e = \left\langle d^\dagger_{xz, Y}d_{xz, Y} - d^\dagger_{yz, X}d_{yz, X} \right \rangle $. We choose $\Phi_e <0$ (opposite in sign to $\Phi_h$). The off-diagonal term $V_{X/Y}$ is defined as
 \begin{align}
V_{X}(k,\theta_X)&=\sqrt{2}v k\sin{\theta_X}\nn\\
&+\frac{p_1}{\sqrt{2}}k^3 \sin{\theta_X}\left(\sin^2{\theta_X}+3 \cos^2{\theta_X})\right )\nn\\
&-\frac{p_2}{\sqrt{2}}k^3 \sin{\theta_X}\cos{2\theta_X},\\
V_Y(k,\theta_Y)&= -V_X(k,\theta_Y).
\end{align}
The band parameters of Eq.~(\ref{electro hamiltonian}) are listed in Table \ref{table:Table S2}. We borrowed the numbers from Refs.~[\onlinecite{fernandes2016low},\onlinecite{fernandes2014distinguishing}], where these parameters have been extracted  from ARPES data.
\begin{table}[b]
\centering
\begin{tabular}{|c|c|c|c|c|c|c|c|c|}
\hline
  $\mu_1$ & $\mu_3$ & $(2m_1)^{-1}$ & $(2m_3)^{-1}$ & a1 & a3 & v & p1 & p2 \\
  \hline
   19.9  & 39.4  & 1.4 & 186 & 136 & -403 & -122 & -137 & -11.7  \\
   \hline
   \end{tabular}
   \caption{Band Parameters for the electron pocket}
     \label{table:Table S2}
     \end{table}
\begin{figure}[t]
		\centering
		\includegraphics[width=1\linewidth]{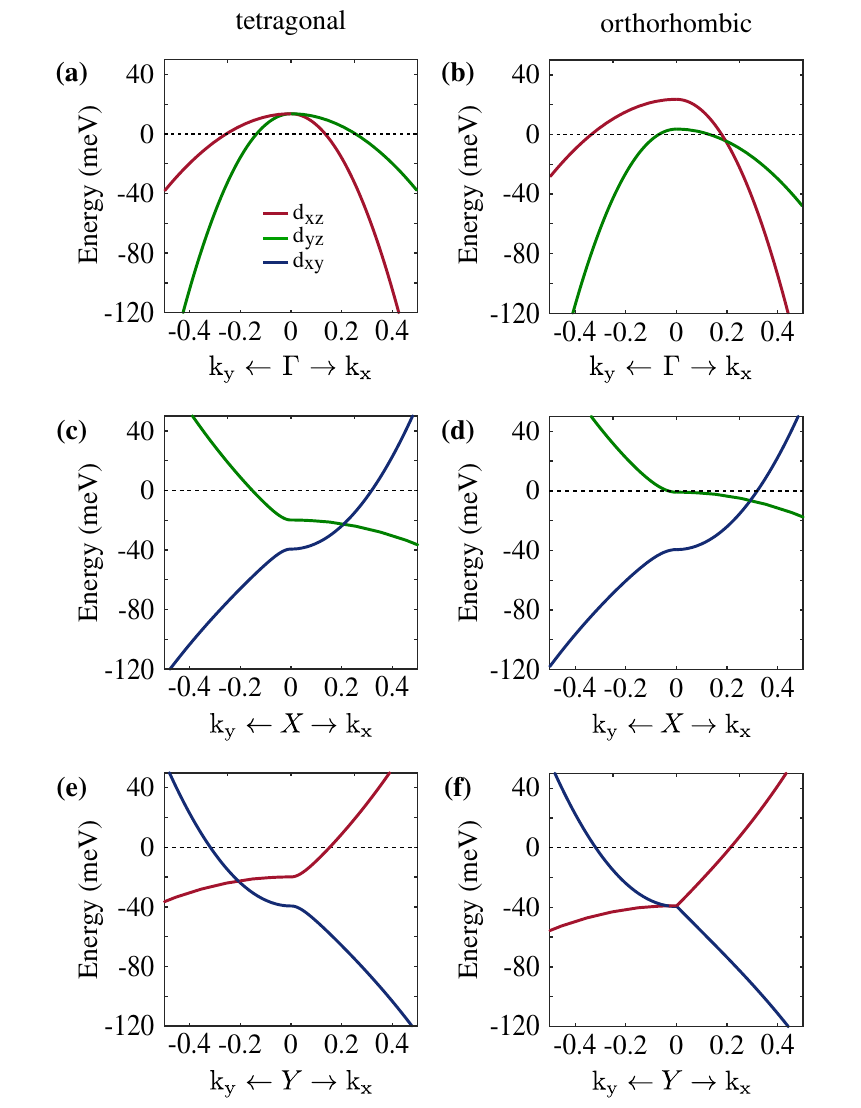}
		\caption{Scenario A: Calculated band dispersion of the 1-Fe unit cell in tetragonal and orthorhombic phase, respectively, near (a-b) $\Gamma$-, (c-d) $X$- and (e-f) $Y$- points of the BZ, respectively. Fitting parameters are taken from Refs.~[\onlinecite{fernandes2016low},\onlinecite{fernandes2014distinguishing}] }
		\label{Fig:1_1A}
\end{figure}

Diagonalizing  Eq.~(\ref{electro hamiltonian}) near the X point, we find that there is a single band that crosses the Fermi level  in both the tetragonal and the orthorhombic phase, see Fig.~\ref{Fig:1_1A}(c,d). The same holds near $Y$, Fig.~\ref{Fig:1_1A}(e,f).  We only consider these bands and neglect the ones which are located fully below $E_F$. The dispersions of the two relevant bands are
\begin{equation}
\xi_{X/Y}=\frac{A_{X/Y}^{(1)}+A_{X/Y}^{(2)}}{2}+\sqrt{\left(\frac{A_{X/Y}^{(1)}-A_{X/Y}^{(2)}}{2}\right)^2+V_{X/Y}^2},
\end{equation}
and the band operators $e_X$ and $e_Y$, in terms of which $H_{X/Y}=\sum_{\kk,\sigma} \xi_{X/Y}(\kk) e_{X/Y,\kk,\sigma}^{\dagger}e_{X/Y,\kk,\sigma}$, are
\begin{align}
e_X=&-i \cos{\phi_X}d_{yz}+\sin{\phi_X}d_{xy},\\
e_Y=&i \cos{\phi_Y}d_{xz}+\sin{\phi_Y}d_{xy},
\end{align}
where
\begin{align}
\cos^2(\phi_{X/Y})=\frac{1}{2}\left[1+\frac{\frac{A_{X/Y}^{(1)}-A_{X/Y}^{(2)}}{2}}{\sqrt{\left(\frac{A_{X/Y}^{(1)}-
A_{X/Y}^{(2)}}{2}\right)^{2}+V_{X/Y}^2}}\right].
\end{align}
\begin{figure}[t]
\includegraphics[width=\columnwidth]{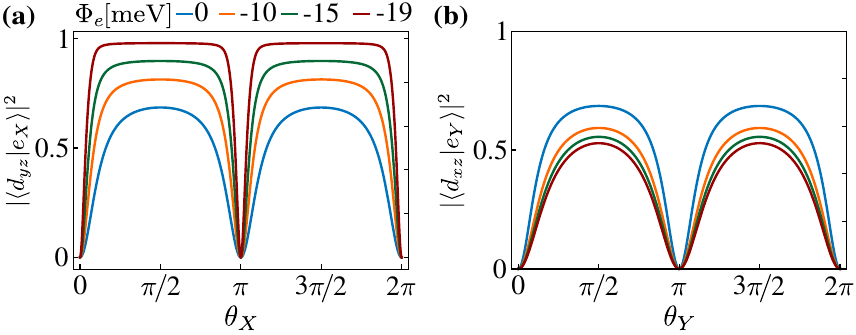}
\caption{The angular variation of orbital contents $|\langle d_{yz}|e_{X}\rangle|^2=\cos^2\phi_X$ on the X-pocket (a), and $|\langle d_{xz}|e_{Y}\rangle|^2=\cos^2\phi_Y$ on the Y-pocket (b) for a set of $\Phi_e$.}
\label{cosphielectron}
\end{figure}
The angular variation of  the orbital $d_{yz/xz}$ content, $|\langle d_{yz/xz}|e_{X/Y} \rangle|^2=\cos^2{\phi_{X/Y}}$  on the Fermi surface is plotted in Fig.~\ref{cosphielectron}. Because of $C_4$ symmetry in the tetragonal phase, $X$ and $Y$ pockets have the same amount of $d_{yz}$ and $d_{xz}$ orbital content (blue lines in Fig.~\ref{cosphielectron}). With increasing $\Phi_e$, the $X$ pocket becomes more of $d_{yz}$ character and deforms into a peanut, while the $Y$ pocket becomes more of $d_{xy}$ character as its $d_{xz}$ content decreases. For our band parameters, $X$ pocket splits into two smaller pocket once $|\Phi_e| \geq 19.9$ meV (the short axis of the peanut becomes zero). Below we limit  $\Phi_e$ to be smaller than this value.

\section{Superconductivity\label{sec:pairing}}

\subsection{Pairing Interaction\label{pairinginteraction}}

The pairing interaction for the model with local fermion-fermion interaction in the band basis has been discussed previously~[\onlinecite{kang2018superconductivity},\onlinecite{kang2018time}]. We include the following components of the interaction Hamiltonian, relevant to the pairing: intra-orbital density-density interaction between fermions on hole and electron pockets, $U_{he}$,  and inter-orbital pair-hopping interaction between fermions on hole and electron pockets,  $J_{he}$, and between the two electron pockets, $J_{ee}$. There are other pairing interactions, i.e., a repulsion within each pocket, but we restrict our consideration to these three as they are enhanced by magnetic fluctuations with momenta $(0,\pi)$, $(\pi,0)$, and $(\pi,\pi)$. The interaction Hamiltonian reads
\begin{align}
&H_{\text{int}}=U_{he}\sum_{\kk,\kk^\prime,\mu}d^\dagger_{\mu,\kk,\uparrow}d^\dagger_{\mu,-\kk,\downarrow}d_{\mu,-\kk^\prime+\Q_\mu,\downarrow}d_{\mu,\kk^\prime+\Q_\mu,\uparrow}\nn\\
+&J_{he}\sum_{\kk,\kk^\prime,\mu\neq\nu}d^\dagger_{\mu,\kk,\uparrow}d^\dagger_{\mu,-\kk,\downarrow}d_{\nu,-\kk^\prime+\Q_\nu,\downarrow}d_{\nu,\kk^\prime+\Q_\nu,\uparrow}\nn\\
+&J_{ee}\sum_{\kk,\kk^\prime,\mu\neq\nu}d^\dagger_{\mu,\kk+\Q_\mu,\uparrow}d^\dagger_{\mu,-\kk+\Q_\mu,\downarrow}d_{\nu,-\kk^\prime+\Q_\nu,\downarrow}d_{\nu,\kk^\prime+\Q_\nu,\uparrow}\nn\\
&+h.c.\label{Eq:InteractionHamiltonian}
\end{align}

We consider only the pairing interaction involving $d_{xz}$ and $d_{yz}$ fermions, i.e., assume that  $\mu,\nu\in\{xz,yz\}$, and $\Q_{xz}=\left(0,\pi\right)$, $\Q_{yz}=\left(\pi,0\right)$. The restriction to $d_{xz}$ and $d_{yz}$ orbitals is justified as $d_{xy}$-fermions have a larger mass~[\onlinecite{chichinadze2019specific}]. To convert the interaction Hamiltonian, Eq.~(\ref{Eq:InteractionHamiltonian}) from the orbital to the band basis we use
\begin{figure}[t]
	\includegraphics[width=\columnwidth]{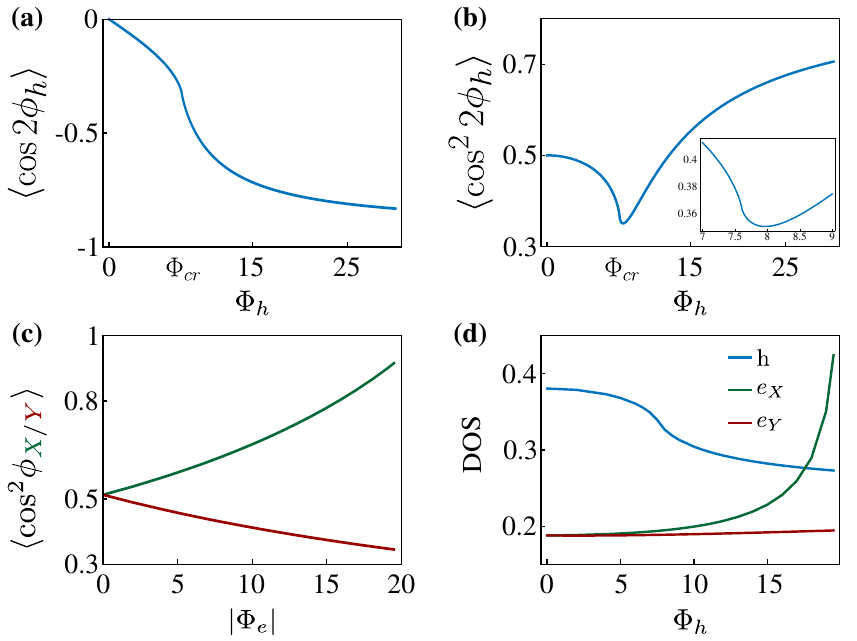}
	\caption{Variation of the angular average of (a) $\langle \cos{2\phi_h}\rangle$, (b) $\langle \cos^2{2\phi_h}\rangle$, (c) $\langle \cos^2{\phi_{X}}\rangle$ and  $\langle\cos^2{\phi_{Y}}\rangle$ on the hole and the electron  pockets with the nematic order $\Phi_{h,e}$. The inset in (b) shows the zoom-in view of $\langle\cos^2{2\phi_h}\rangle$ near $\Phi_{cr}$. The non-analyticity is at $\Phi_{cr} \approx 7.6$ meV. (d) Variations of the density of state (DOS) on different pockets with $\Phi_{h}$. }
	\label{DOS}
\end{figure}
\begin{align}
 d_{xz,\kk}&=\sin{\phi_h(\kk)} h_{\kk},\hspace{.2cm} d_{xz,\kk+\Q_{xz}}=\cos{\phi_Y(\kk) e_{Y,\kk}}, \\ d_{yz,\kk}&=\cos{\phi_h(\kk)h_{\kk},\hspace{.2cm} d_{yz,\kk+\Q_{yz}}=\cos{\phi_{X}(\kk)e_{X,\kk}}}.
\label{transformations}
\end{align}
Substituting these into Eq.~(\ref{Eq:InteractionHamiltonian}), we obtain the pairing interaction in the band basis
\begin{align}
	&H_\text{pair}=\sum_{\kk,\pp}h_{\kk,\uparrow}^{\dagger}h_{-\kk,\downarrow}^{\dagger}\times\nn\\
	&\Bigg[U_s \left( e_{X,-\pp,\downarrow}e_{X,\pp,\uparrow} \cos^2{\phi_X} +e_{Y,-\pp,\downarrow}e_{Y,\pp,\uparrow}  \cos^2{\phi_Y}\right)+\nn\\
	&U_d \cos{2\phi_h}\left(e_{X,-\pp,\downarrow}e_{X,\pp,\uparrow} \cos^2{\phi_X} -e_{Y,-\pp,\downarrow}e_{Y,\pp,\uparrow} \cos^2{\phi_Y}\right)\Bigg]\nn\\
	&+  J_{ee} \cos^2{\phi_X} \cos^2{\phi_Y}  e_{X,\kk,\uparrow}^{\dagger}e_{X,-\kk,\downarrow}^{\dagger}e_{Y,-\pp,\downarrow}e_{Y,\pp,\uparrow}+h.c,
	\label{Pairing Hamiltonian}
\end{align}
where, $U_s=\frac{U_{he}+J_{he}}{2}$ and $U_d=\frac{U_{he}-J_{he}}{2}$ are $s$- and $d$-wave components of the pairing interaction between the hole and the electron pockets. We use $\alpha=\frac{U_d}{U_s}$ to measure relative strength of this part of the interaction in the $s$-wave and the $d$-wave channels.

\subsection{Gap Equation\label{gapequation}}
	
We introduce the gap functions $\Delta_h$ on the hole pocket and $\Delta_X$ and $\Delta_Y$ on the electron pockets. The equations for $\Delta_h$, $\Delta_X$, and $\Delta_Y$  are obtained by solving $3\times 3$ matrix equation. We present the BCS gap equations in Appendix~(\ref{Appendix:GapEquations}), Eqs. (\ref{hole gap equation}-\ref{Y gap equations}).  The solution of these gap equations is
\begin{align}
\Delta_h&=\Delta_1+\Delta_2 \cos{2\phi_h},\\
\label{gapform}
\Delta_X&=\Delta_3 \cos^2{\phi_X},\\
\Delta_Y&=\Delta_4 \cos^2{\phi_Y}.
\end{align}
 At $T\approx T_c$, $\Delta_i(i=1,\cdots 4)$ are the solutions of the matrix equation:
\begin{widetext}
\begin{equation}
 \small  \begin{bmatrix}
\Delta_1 \\ \Delta_2 \\ \Delta_3 \\ \Delta_4
\end{bmatrix}=\frac{1}{\lambda}\begin{bmatrix}
0 & 0 & -N_X \langle \cos^4{\phi_X} \rangle  &  -N_Y \langle \cos^4{\phi_Y}\rangle\\  0 & 0 & -N_X \alpha \langle \cos^4{\phi_X}\rangle & N_Y \alpha \langle \cos^4{\phi_Y}\rangle  \\
-N_h \langle 1+\alpha \cos{2\phi_h}\rangle & -N_h \langle  \cos{2\phi_h}+\alpha \cos^2{2\phi_h}\rangle & 0 & -\frac{J_{ee}}{U_s} N_Y \langle \cos^4{\phi_Y}\rangle \\
-N_h \langle 1-\alpha \cos{2\phi_h}\rangle & -N_h \langle  \cos{2\phi_h}-\alpha \cos^2{2\phi_h}\rangle & -\frac{J_{ee}}{U_s}N_X \langle \cos^4{\phi_X}\rangle & 0
\end{bmatrix}\begin{bmatrix}
\Delta_1 \\ \Delta_2 \\ \Delta_3 \\ \Delta_4
\end{bmatrix}.
\label{linearized gap equation full}
\end{equation}
\end{widetext}
Here $\lambda$ is the eigenvalue of the gap matrix defined as $
\frac{1}{\lambda}=U_s \log(\frac{\Lambda}{T})$, $\langle A \rangle$ defines the angular average of $A$ over the corresponding Fermi surface pocket, and $N_X$, $N_Y$ and $N_h$ are the densities of states for the $X$, $Y$, and the $\Gamma$ pocket, respectively. In Fig.~\ref{DOS}(a,b,c), we show the variation of $\langle\cos{2\phi_h}\rangle$, $\langle\cos^2{2\phi_h}\rangle$, $\langle\cos^4{\phi_X}\rangle$, and $\langle\cos^4{\phi_Y}\rangle$ as a function of the nematic order $\Phi_{h,e}$. We find that $\langle\cos{2\phi_h}\rangle$ and $\langle\cos^2{2\phi_h}\rangle$  exhibit a kink like non analyticity near $\Phi_h=\Phi_{cr}$. In the Appendix we show that the singularities (non-analyticities)  are $x \log(x)$, and $x^2 \log(x)$, where $x=\Phi_h/\Phi_{cr} -1$. The densities of states also depend on $\Phi_{h,e}$, as we show them in the Fig.~\ref{DOS}(d).

\begin{figure}[t]
\includegraphics[width=\columnwidth]{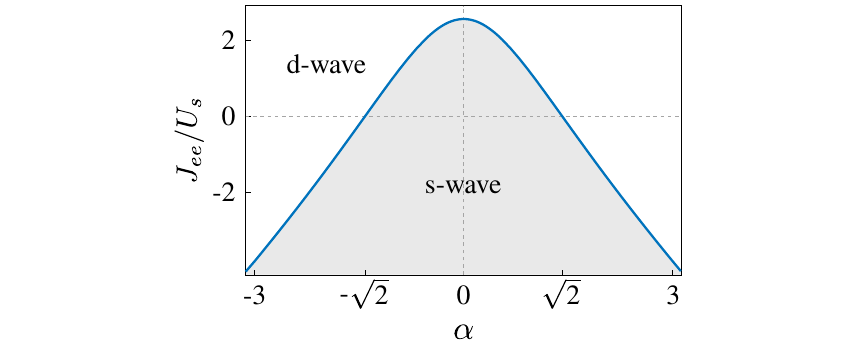}
\caption{Regions of $s$-wave and $d$-wave superconductivity according to the solution of Eq.~(\ref{non jee}) for different $\alpha$ and
  $J_{ee}$ in the absence of nematicity.}
\label{jeenotzero}
\end{figure}

We numerically solve Eq.~(\ref{linearized gap equation full}) and obtain $T_c$ and find the gap structure $\Delta=\left(\Delta_1,\Delta_2,\Delta_3,\Delta_4\right)$ for the leading superconducting instability.
In the tetragonal phase, $N_X=N_Y=N_e$ and $\langle  \cos{2\phi_h}\rangle = \langle \cos{2\theta_h}\rangle =0$. Then $s^{\pm}-$wave and $d$-wave pairing channels are decoupled.
 The eigenvalues of the gap matrix, Eq(\ref{linearized gap equation full}), are
\begin{eqnarray}
\lambda_s=\lambda_0 \left[-\frac{J_{ee}}{Us}+\sqrt{\left(\frac{J_{ee}}{U_s}\right)^2+8 \frac{N_h}{N_e \langle \cos^4{\phi_X}\rangle}}\right],\\
\lambda_d=\lambda_0 \left[\frac{J_{ee}}{Us}+\sqrt{\left(\frac{J_{ee}}{U_s}\right)^2+4 \frac{N_h}{N_e \langle \cos^4{\phi_X}\rangle} \alpha^2}\right],
\end{eqnarray}
where $\lambda_0=\frac{N_e \langle \cos^4{\phi_X}\rangle}{2}$. For $J_{ee}=0$, the gap function is either $s$-wave, for $\alpha<\sqrt{2}$, or $d$-wave, for $\alpha>\sqrt{2}$. For $J_{ee}\neq 0$, superconductivity is s-wave when
\begin{equation}
 \frac{J_{ee}}{U_s}< \sqrt{\frac{N_h}{2 N_e \langle \cos^4{\phi_X}\rangle }}\frac{(2-\alpha^2)}{\sqrt{\alpha^2+2}}.
 \label{non jee}
 \end{equation}
The phase diagram for Eq.~\ref{non jee} is shown in Fig.~\ref{jeenotzero}.
We next move to the nematic phase. Now $\langle  \cos{2\phi_h}\rangle \neq 0$, and both $\Delta_1$ and $\Delta_2$ are non-zero for any $\alpha$ and $J_{ee}$.

To simplify the presentation, we neglect $J_{ee}$. Without nematicity,  superconducting  order is $s$-wave for $\alpha < \sqrt{2}$ and $\Delta_h = \Delta_1$. At small $\Phi$ (i.e., small $\Phi_h$ and $\Phi_e$), $\phi_h \approx \theta_h$ and $\Delta_2 \propto \Phi$. This gives rise to $\Phi \cos{2\theta_h}$ angular variation of $\Delta_h$. If this was the only effect of nematicity, the angle variation would grow with $\Phi$, and $\Delta_h$ would necessary develop a deep minima and then gap nodes. However, as $\Phi$ increases, $\phi_h$ deviates from $\theta_h$ due to orbital transmutation, and at large $\Phi$ becomes $\pi/2$  almost everywhere on the hole pocket. Then $\Delta_2 \cos{2\phi_h}$ term becomes angle-independent, and the gap function on the hole pocket recovers a pure $s$-wave form.  Besides, due to the same orbital transmutation, the magnitude $\Delta_2$ becomes a non-linear function of $\Phi$ and not necessary exceeds $\Delta_1$ even at large $\Phi$.

A similar situation holds if $\alpha > \sqrt{2}$, when the superconducting order without nematicity is $d$-wave, $\Delta_h = \Delta_2 \cos {2 \theta_h}$. At a small $\Phi$, the key effect of nematicity is an admixture of $\Delta_1$.  At large $\Phi$, $\theta_h \to \phi_h \approx  \pi/2$, and the  nodes disappear.

\begin{figure}[t]
  \includegraphics[width=\columnwidth]{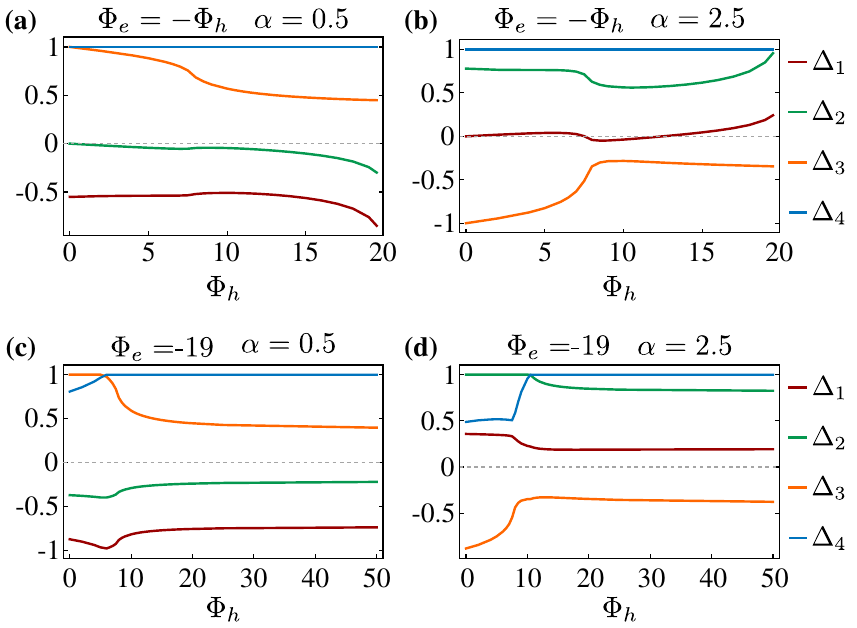}
 \caption{ Variations of the gap amplitudes\\ $\Delta=\left(\Delta_1,\Delta_2,\Delta_3,\Delta_4\right)$ with the nematic order $\Phi_{e,h}$ and interaction ratio $\alpha$ for: (a) $\alpha=0.5$, $\Phi_h=-\Phi_e$, (b) $\alpha=2.5$, $\Phi_h=-\Phi_e$, (c) $\alpha=0.5$, $\Phi_e=-19$ meV, (d) $\alpha=2.5$, $\Phi_e=-19$ meV.}
\label{gap structure at small and large alpha}
\end{figure}

\begin{figure}[t]
\includegraphics[width=\columnwidth]{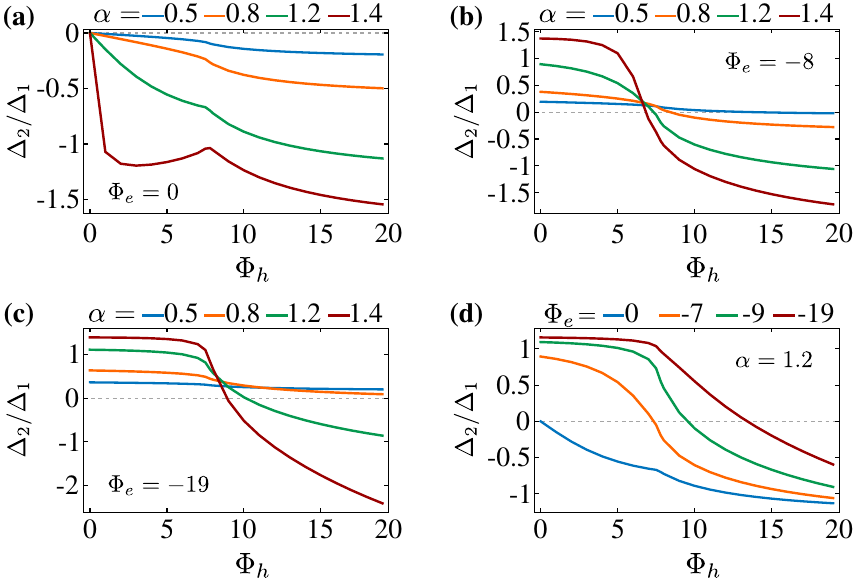}
   \caption{ The variation of $\Delta_2/\Delta_1$ with $\Phi_h$ for different interaction ratios $\alpha$ for a fixed electron nematic order (a) $\Phi_e=0$ meV, (b) $\Phi_e=-8$ meV,and (c)$\Phi_e=-19$ meV. In (d) we fix $ \alpha=1.2$ and plot $\Delta_2/\Delta_1$ with $\Phi_h$ for different values of electron nematic order $\Phi_e$.}
\label{ratio with phih}
\end{figure}

\begin{figure}[t]
\includegraphics[width=\columnwidth]{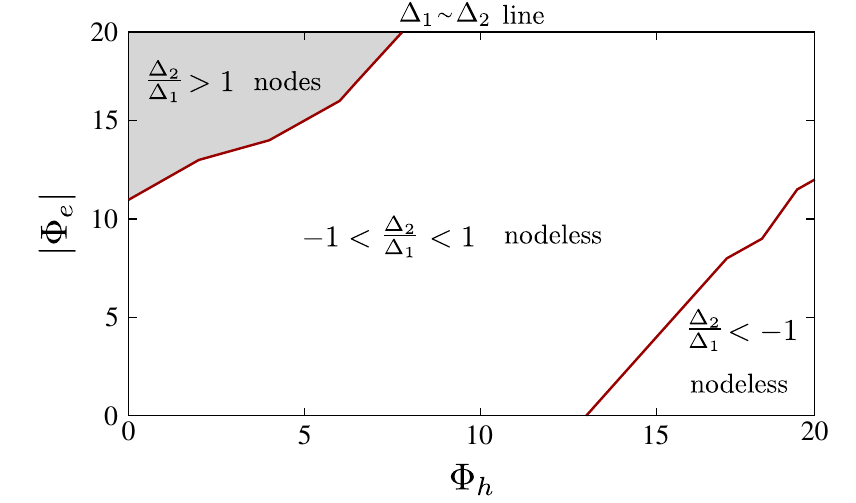}
\caption{The variation of $\Delta_2/\Delta_1$ as a function of $\Phi_h$ and $\Phi_e$. We set $\alpha=1.2$ in this plot.}
\label{ratio in phih phie space}
\end{figure}

\begin{figure}[t]
\includegraphics[width=\columnwidth]{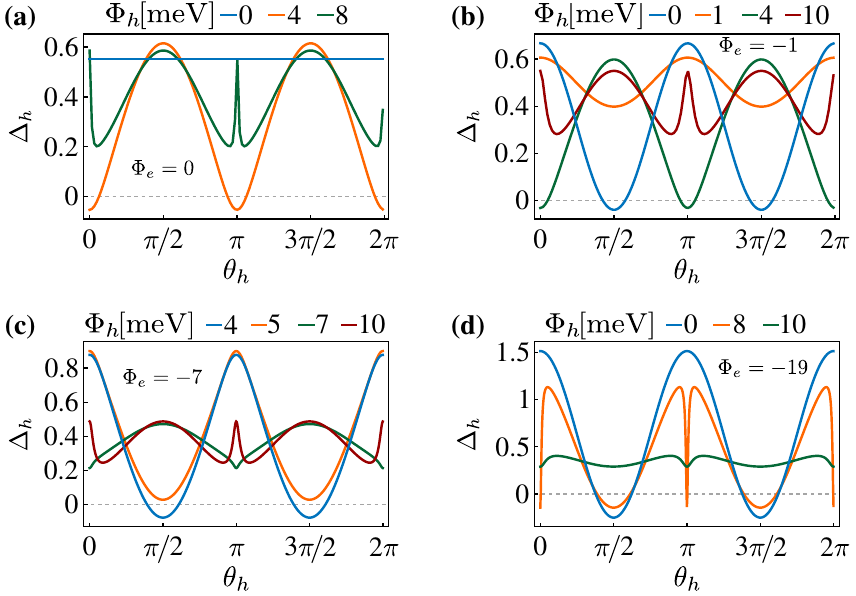}
   \caption{The angular variation of $\Delta_h(\theta_h)$ with $\theta_h$ for various values of $\Phi_h$ at (a) $\Phi_e=0$ meV, (b)$\Phi_e=-1$ meV, (c) $\Phi_e=-7$ meV, and (d)$\Phi_e=-19$ meV. We set $\alpha=1.4$ }
\label{evolution of nodes}
\end{figure}

\begin{figure}[t]
\includegraphics[width=\columnwidth]{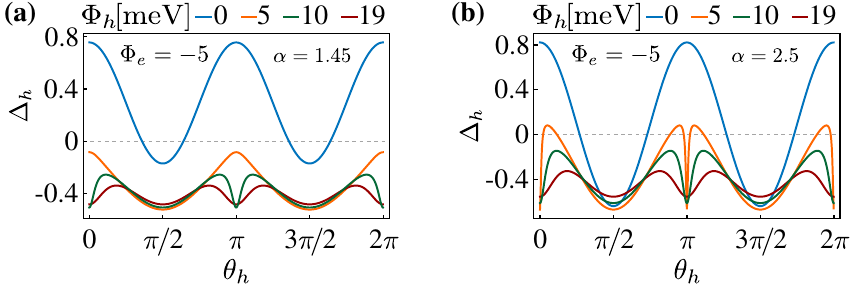}
\caption{Angular variation of $\Delta_h(\theta_h)$ at selected values of $\Phi_h$ for (a) $\alpha=1.45$ and (b) $\alpha=2.5$. We set $\Phi_e=-5$ meV. }
\label{node disappear in d}
\end{figure}

The questions, which we address below  are (i) whether for $\alpha < \sqrt{2}$ the nodes in $\Delta_h$ develop at intermediate $\Phi_h$ and (ii) how the nodes in $\Delta_h$ disappear for $\alpha > \sqrt{2}$ as $\Phi_h$ increases. To address these issues we solve the gap equations for different $\alpha$ at various $\Phi_h$ and $\Phi_h/\Phi_e$.  We show the results in Figs.~\ref{gap structure at small and large alpha}-\ref{node disappear in d}.
Before we discuss these results, several general observations are in order. According to Eq.~(\ref{gapform}), $\Delta_h$ has a node at an angle $\theta_0$ if
\begin{equation} \cos{2\phi_h}(\theta_0,\Phi_h)=-\dfrac{\Delta_1}{\Delta_2}.
\label{node condition}
\end{equation}
The ratio $\Delta_2/\Delta_1$ depends on $\Phi_h,\Phi_e$, and on $\alpha$. Obviously, the nodes are possible only if $|\Delta_2/\Delta_1| >1$.  Shrinking the angular variation of $\cos{2\phi_h}$ at $\Phi_h > \Phi_{cr}$ puts additional restriction on $\Delta_2/\Delta_1$ for the nodes to appear. Further, the number of possible nodes changes between $\Phi_h<\Phi_{cr}$ and $\Phi_h > \Phi_{cr}$. In the first case,  the gap functions at $\theta=0$ and $\dfrac{\pi}{2}$ are $\Delta_1+\Delta_2$ and $\Delta_1-\Delta_2$, respectively. When $\left|\Delta_2/\Delta_1\right|>1$, the two have opposite signs, hence there have to be an odd number of nodes between $0$ and $\tfrac{\pi}{2}$;  the total number of nodes  is then $4,12,20, \dots$. For $\Phi_h >\Phi_{cr}$,  $\Delta_h(\theta)$ at $\theta=0$ and $\pi/2$ become the same $\Delta_1-\Delta_2$ due to orbital transmutation. Then, there have to be an even number of nodes between $0$ and $\pi/2$, hence the total number of nodes is $0,8,16, \dots$.

In our case,  we find (see Appendix for details)
 \begin{eqnarray}
\dfrac{\Delta_2}{\Delta_1}&=& 2 \alpha \dfrac{g+\alpha \langle \cos{2\phi_h}\rangle }{\left(1-\alpha^2  \langle \cos^2{2\phi_h}\rangle \right)+D},
\label{d1 d2 ratio}
\end{eqnarray}
where
\begin{eqnarray}
g = g(\Phi_e)&=&  \dfrac{N_X \langle \cos^4{\phi_X}\rangle- N_Y \langle \cos^4{\phi_Y}\rangle}{N_X\langle \cos^4{\phi_X}\rangle + N_Y \langle \cos^4{\phi_Y}\rangle},
\end{eqnarray}
and
\begin{align}
D= &\Big[4 \alpha^2 \left(\langle \cos{2\phi_h}\rangle^2-\langle \cos^2{2\phi_h}\rangle\right)\left(1-g^2\right)\nn\\
&+\left(1 +2 g \alpha \langle \cos{2\phi_h}\rangle +\alpha^2 \langle \cos^2{2\phi_h}\rangle\right)^2\Big]^{1/2}.
\end{align}
The function  $g(\Phi_e)$ measures the asymmetry between X and Y pockets at a non-zero $\Phi_e$. We find that it increases roughly  linearly with $|\Phi_e|$.\newline

We now discuss the results.
In Fig.~\ref{gap structure at small and large alpha}(a,c) we show $\Delta_i$ for $\alpha =0.5$ and in (b,d) for $\alpha =2.5$, when the primary order is $s$-wave and $d$-wave, respectively.  We see that for $\alpha =0.5$, the magnitude of the s-wave component $\Delta_1$ far exceeds $\Delta_2$  of  the $d$-wave component, i.e., the gap remains an $s$-wave with a small admixture of $d$-wave.  For $\alpha =2.5$, the situation is opposite -- the gap remains predominantly $d$-wave with a small admixture of an $s$-wave.  In both  cases therefore, the effect of nematicity is rather weak, even when $\Phi_h$ is large.

In Fig.~(\ref{ratio with phih}) we plot $\Delta_2/\Delta_1$ as a function of $\Phi_h$ for various $\Phi_e$ and $\alpha < \sqrt{2}$. We see that  when $\alpha$ is not close to $\sqrt{2}$, then $|\Delta_2/\Delta_1|<1$ for any $\Phi_h$ and $\Phi_e$.  As a consequence, there are no nodes in the gap function. This agrees with Fig.~\ref{gap structure at small and large alpha}. However, for $\alpha \leq \sqrt{2}$, we find intervals of $\Phi_h < \Phi_{cr}$, where $|\Delta_2/\Delta_1|>1$.
This holds, e.g.,  for $\alpha =1.4$ and $\Phi_e =0$ (dark red curve in Fig.~\ref{ratio with phih}(a)).  By our generic reasoning, there must be 4 nodes. The same holds for the same $\alpha$ and  sizable $\Phi_e$
(see Figs. \ref{ratio with phih}(b-c)). The only difference is that for $\Phi_e=0$, the 4 nodes are near $k_x$-direction, while for sizable $\Phi_e$ they are near  $k_y$-direction.

Next, we see from Fig.~\ref{ratio with phih} that the ratio $|\Delta_2/\Delta_1|$ evolves around $\Phi_h = \Phi_{cr}$ and even changes sign for sizable $\Phi_e$.  For larger $\Phi_h$ we again have $|\Delta_2/\Delta_1|>1$ for $\alpha \leq \sqrt{2}$.  However,  this no longer guarantees  the existence of the nodes as by our general reasoning above their number can be zero. We will see that this is what happens -- the nodes do not develop despite $|\Delta_2/\Delta_1| >1$ because $\phi_h$ clusters around $\pi/2$.

In Fig.~\ref{ratio in phih phie space} we mark the boundaries of $|\Delta_2/\Delta_1|$ on the ($\Phi_h,\Phi_e$) plane at a fixed $\alpha =1.2$.  The area of the "corner" regions, where $|\Delta_2/\Delta_1| >1$, increases when $\alpha$ approaches $\sqrt{2}$.  As we mentioned, the nodes only develop in the left upper grayish colored corner, where $\Phi_h < \Phi_{cr}$.
In  Fig.~\ref{evolution of nodes} we plot the gap function $\Delta_h(\theta_h)$. We find
$4$ different scenarios how nodes can appear/disappear when one varies $\Phi_h$  at a fixed value of $\Phi_e$ and $\alpha$ slightly below critical $\sqrt{2}$. Here, we further set  $\alpha=1.4$.
\begin{enumerate}
\item In Fig.~\ref{evolution of nodes}(a) we set  $\Phi_e =0$. There are no nodes at $\Phi_h=0$ in agreement with
Fig.~\ref{ratio with phih}(a).  At  $\Phi_h \approx 1$ meV, 4 nodes appear near the $k_x$-direction. They exist up to $\Phi_h \leq \Phi_{cr}$ and disappear at larger $\Phi_h$. In the node count, the number of nodes changes  with $\Phi_h$ as  $0 \rightarrow 4 \rightarrow 0$.
\item  In Fig.~\ref{evolution of nodes}(b)  we set $\Phi_e = -1$ meV. In this case there are $4$ nodes near $k_y$-direction already for $\Phi_h=0$. As $\Phi_h$ increases, the 4 nodes disappear  at $\Phi_h \sim 1$ meV due to non-monotonic behavior of $\Delta_2/\Delta_1$, like in Fig.~\ref{ratio with phih}(b,c). As $\Phi_h$ increases further,  4 nodes re-appear, now near $k_x$-direction, at  $\Phi_h \sim 3$ meV. These nodes then disappear at $\Phi_h \leq \Phi_{cr}$. In this case, the number of nodes changes with $\Phi_h$ as  $4 \rightarrow 0 \rightarrow 4\rightarrow 0$.
\item In Fig.~\ref{evolution of nodes}(c) we set  $\Phi_e =-7$ meV. In this case,  at small $\Phi_h$ there are $4$ nodes near  $k_y$-direction.  These nodes disappear at some $\Phi_h \leq \Phi_{cr}$. In this case, number of nodes changes with $\Phi_h$ as  $4 \rightarrow 0$
\item In Fig.~\ref{evolution of nodes}(d) we set $\Phi_e -19$ meV. In this case, there are 4 nodes  near $k_y$-direction for all $\Phi_h \leq \Phi_{cr}$. For $\Phi_h> \Phi_{cr}$,  the number of nodes first increases from $4$ to $8$, because the gap function along the $k_x$- and the $k_y$-direction becomes nearly the same and has to cross zero twice.  As $\Phi_h$ increases further, the 8 nodes disappear due to clustering of $\phi_h$ near $\pi/2$. In this case, the number of nodes changes with $\Phi_h$ as $4 \rightarrow 8 \rightarrow 0$.
\end{enumerate}
For $\alpha>\sqrt{2}$, superconducting order  in the tetragonal phase is d-wave with $4$ nodes on the hole pocket. With increasing nematic order the nodes  disappear due to orbital transmutation  either because $\Delta_1$ becomes larger than $\Delta_2$  or $\Delta_2$ remains larger than $\Delta_1$, but $\phi_h$ clusters around $\pi/2$. In Fig.~\ref{node disappear in d} we  show the results for $\Delta_h$ at two values of $\alpha > \sqrt{2}$. For $\alpha=1.45$, the nodes  $\Delta_h$ disappear because $\Delta_1$ becomes larger than $\Delta_2$. This happens  at  $\Phi_h < \Phi_{cr}$, i.e., well before $\phi_h$ starts clustering near $\pi/2$.  In this case, the number of nodes changes with $\Phi_h$ as $4\rightarrow 0$. For $\alpha=2.5$ $\Delta_2$ remains larger than $\Delta_1$, and the nodes disappear at $\Phi_h >  \Phi_{cr}$ due to clustering of $\phi_h$. We see from the Figure that in this case the number of nodes changes with $\Phi_h$ as $4 \rightarrow 8 \rightarrow 0$. (4 on a blue line, 8 on a  orange, and 0 on green and red  lines).  A nodeless gap deep in orthorhombic phase for $\alpha > \sqrt{2}$ is consistent with RPA calculations of Ref.~[\onlinecite{rhodes2020non}]. \newline

Note that the results for a non-zero $J_{ee}$ are quite similar, only the value of $\alpha$ near which the system develops nodes coming out of an $s$-wave superconductor at $\Phi =0$ shifts from  $\alpha =\sqrt{2}$.

\subsection{Temperature dependence of gap\label{sec:gap_T}}

In this section we obtain the temperature dependence of $  \Delta_i(T) $ near the superconducting transition. We will use the result for $  \Delta_i(T) $ in the next section, where we compute the jump of the specific heat at $T_c$. We assume that the ratios $\Delta_i/\Delta_j$ do not change substantially with temperature and parametrize four gap functions as
\begin{equation}
 \Delta(T)=\Delta_0(T) \left(\Delta_1,\Delta_2,\Delta_3,\Delta_4\right)=\Delta_0(T) \mathbf{\Delta},
 \end{equation}
where $\mathbf{\Delta}=\left(\Delta_1,\Delta_2,\Delta_3,\Delta_4\right)$  are the same (up to an overall factor) as we obtained in  Sec.~\ref{gapequation}  by solving the linearized gap equations (\ref{linearized gap equation full}). We normalize $\mathbf{\Delta}$ by setting its largest component equal to $1$. To simplify the presentation, we again first assume $J_{ee}=0$ and then  present the results  for a non-zero $J_{ee}$.

The non-linear equation  for the gap on the hole pocket is
\begin{align}
\Delta_1&+\Delta_2 \cos{2\phi_h}\nn\\
=-\Bigg[&\Delta_3(U_s+U_d \cos{2\phi_h}) \int_{\pp} \frac{\tanh( \frac{E_X}{2T})}{2 E_X}\cos^4{\phi_X}  \nn\\
+&\Delta_4 (U_s-U_d \cos{2\phi_h}) \int_{\pp} \frac{\tanh( \frac{E_Y}{2T})}{2 E_Y}\cos^4{\phi_Y}    \Bigg],
\label{full gap on hole}
\end{align}
where $E_X = \sqrt{\xi^2_x + \Delta^2_3 \Delta^2_0 (T)}, ~E_Y = \sqrt{\xi^2_y + \Delta^2_4 \Delta^2_0 (T)}$. Multiplying Eq.~(\ref{full gap on hole}) by $\Delta_1+\Delta_2 \cos{2\phi_h}$,  averaging over the hole Fermi surface pocket, and expanding the r.h.s. to order $\Delta^2_0 (T)$ as
 \begin{align}
&\int_{\pp} \frac{\tanh(\frac{E_{X/Y}}{2T})}{2 E_{X/Y}}\cos^4{\phi_{X/Y}(\pp)}=\nn\\
&N_{X/Y}\left(\log\frac{\Lambda}{T}\left\langle\cos^4{\phi_{X/Y}} \right \rangle  -K \Delta_3^2 \frac{\Delta_0^2}{T_c^2}\langle\cos^8{\phi_{X/Y}} \rangle \right)+O(\Delta_0^4),
\label{eq2}
\end{align}
where $K=\frac{7 \zeta(3)}{8 \pi^2}$, we obtain
\begin{align}
&N_h \log\frac{\Lambda}{T_c} \left\langle \left(\Delta_1+\Delta_2 \cos{2\phi_h}\right)^2\right\rangle =\nn\\
&\log\frac{\Lambda}{T}\left[N_x \Delta_3^2 \left\langle \cos^4{\phi_X}\right\rangle  + N_Y \Delta_4^2 \langle \cos^4{\phi_Y}\rangle \right]\nn\\
&-K \frac{\Delta_0(T)^2}{T_c^2} \left[N_x \Delta_3^4 \langle \cos^8{\phi_X}\rangle + N_Y \Delta_4^4 \langle \cos^8{\phi_Y}\rangle \right].
\label{eq4}
\end{align}
Multiplying the Eq.~(\ref{hole equation}) by $\Delta_1+\Delta_2 \cos{2\phi_h}$, averaging over the hole Fermi surface pocket, and using Eqs.(\ref{X equation})-(\ref{Y equation}), we obtain the relation
\begin{align}
&N_h\langle (\Delta_1+\Delta_2 \cos{2\phi_h})^2\rangle\nn\\
&=\left(N_X \Delta_3^2 \langle\cos^4{\phi_X} \rangle+ N_Y \Delta_4^2\langle\cos^4{\phi_Y} \rangle\right),
\label{connection between hole and electron}
\end{align}
Approximating  $\log{\frac{\Lambda}{T}} \approx \log\frac{\Lambda}{T_c}+\frac{T_c-T}{T_c}$  and using Eq.~(\ref{connection between hole and electron}), we obtain from (\ref{eq4})
\begin{equation}
\Delta_0(T)^2=\frac{T_c(T_c-T)}{K}\frac{N_X\Delta_3^2 \langle \cos^4{\phi_X}\rangle+N_Y \Delta_4^2 \langle \cos^4{\phi_Y}\rangle}{N_X \Delta_3^4 \langle \cos^8{\phi_X}\rangle+N_Y \Delta_4^4 \langle \cos^8{\phi_Y}\rangle}.
\label{TVariationof gap}
\end{equation}
We recall that $\Delta_3$ and $\Delta_4$ are functions of $\Phi_h$, $\Phi_e$ and $\alpha$.

For $J_{ee}\neq 0$, the same procedure yields
\begin{widetext}
\begin{eqnarray}
\Delta_0(T)^2 =\frac{T_c(T_c-T)}{K}\frac{N_X\Delta_3^2 \langle \cos^4{\phi_X}\rangle+N_Y \Delta_4^2 \langle \cos^4{\phi_Y}\rangle+2 \frac{J_{ee}}{\lambda}N_X N_Y \Delta_3 \Delta_4 \langle \cos^4{\phi_X}\rangle \langle \cos^4{\phi_Y} \rangle }{N_X \Delta_3^4 \langle \cos^8{\phi_X}\rangle + N_Y \Delta_4^4 \langle \cos^8{\phi_Y}\rangle + \frac{J_{ee}}{\lambda}N_X N_Y \Delta_3 \Delta_4\left(\Delta_3^2 \langle \cos^8{\phi_X}\rangle \langle \cos^4{\phi_Y}\rangle + \Delta_4^2 \langle \cos^8{\phi_Y}\rangle \langle \cos^4{\phi_X}\rangle \right) }, \nonumber \\
\end{eqnarray}
\end{widetext}
where $\lambda$ is the largest eigenvalue of Eq.~(\ref{linearized gap equation full}).

\section{Specific Heat \label{sec:spec_heat}}

In this section we examine the specific heat jump at $T_c$ and  its band-resolved composition, as a function of nematicity for scenarios A and B. In the mean-field approximation the specific heat is the sum of contributions from $\Gamma$, X and Y pockets: 
\begin{align}
C_{ v}=\sum_{i=h,X,Y}\int_{\kk} \left(\frac{E_i^2(\kk)}{2T^2}-\frac{1}{4T}\frac{\partial |\Delta_i(\kk)|^2}{\partial T}\right)\frac{1}{\cosh^2\left(\frac{E_i(\kk)}{2T}\right)}.\label{Eq:Specificheat}
\end{align}

The first term in the r.h.s of Eq.~(\ref{Eq:Specificheat}) is the normal state contribution  at $T= T_c+0^+$. Evaluating the k-integral we obtain
\begin{equation}
C_{ v}=\frac{2}{3}\pi^2 T_c(N_h+N_X+N_Y).
\end{equation}
The second term in the r.h.s of Eq.~(\ref{Eq:Specificheat}) accounts for the jump of $\Delta C_v$ at $T_c$.
  It is equal to
\begin{align}
\Delta C_{v} & =  -\frac{1}{4 T_c}\sum_{i=h,X,Y} \int_{\kk} \frac{1}{\cosh(\frac{  \xi_i(\kk)}{2 T_c})^2} \frac{d}{dT}\Delta_{i}(\theta)^2\nn\\
&=-\sum_{i=h,X,Y} N_i \int_0^{2\pi} \frac{d\theta}{2\pi} \frac{d}{dT}\Delta_{i}(\theta)^2.
\label{CV}
\end{align}

Substituting the results for the gap functions, we find that
\begin{align}
\Delta C_{ v} &=-\frac{d}{dT}\Delta_0(T)^2 \Big[N_h \langle \left(\Delta_1+\Delta_2 \cos{2\phi_h}\right)^2 \rangle\nn\\
&+N_X \Delta_3^2 \langle\cos^4{\phi_X}\rangle +N_Y  \Delta_4^2 \langle\cos^4{\phi_Y}\rangle\Big]\nn\\
&= \Delta C^h_v +\Delta C^X_v +\Delta C^Y_v.
\label{cv form}
\end{align}
Setting $J_{ee} =0$ and  using Eq.~(\ref{connection between hole and electron}), we find that  $\Delta C_v^h =\Delta C_v^X +\Delta C_v^Y$. Using Eqs.(\ref{connection between hole and electron},\ref{TVariationof gap},\ref{cv form}), we find
\begin{align}
\frac{\Delta C_v}{C_{v}}&= 1.43  \frac{  2N_h^2 \left\langle \Delta_h(\theta)^2\right\rangle^2}{N \left(N_X \Delta_3^4 \langle \cos^8{\phi_X}\rangle + N_Y \Delta_4^4 \langle \cos^8{\phi_Y}\rangle  \right)},
\label{final expression}
\end{align}
where $1.43$ is the BCS result for a single band superconductor, and $N=N_h+N_X+N_Y$. Without a nematic order, the ratio would be
\begin{equation}
\left(\frac{\Delta C_v}{C_v}\right)_{\Phi=0}=1.43 \frac{2}{1+\frac{N_h}{2 N_e}}\frac{\langle\cos^4{\phi_X}\rangle^2}{\langle\cos^8{\phi_X}\rangle}.
\end{equation}
If the electron pockets  would consist solely of $d_{xz}$ and $d_{yz}$ fermions, we would obtain $\Delta C_v/C_v|_{\Phi=0}=2.86/(1+N_h/(2N_e))$. For the parameters from Tables (\ref{table:Table S1}-\ref{table:Table S2}) this  yields $\Delta C_v/C_v|_{\Phi=0}= 1.42$. In presence of the $d_{xy}$ orbital, however, $\Delta C_v/C_v|_{\Phi=0} \approx 1.09$. The smallness comes from the fact that relatively heavy $d_{xy}$ band contributes to $C_v(T)$ in the normal state, but not to $\Delta C_v$. This is similar to the case of KFe$_2$As$_2$ (Ref~\citep{chichinadze2019specific}).

\subsection{ Specific heat jump at $T_c$ for scenario A \label{sec:scenarioB}}

\begin{figure}[t]
\includegraphics[width=\columnwidth]{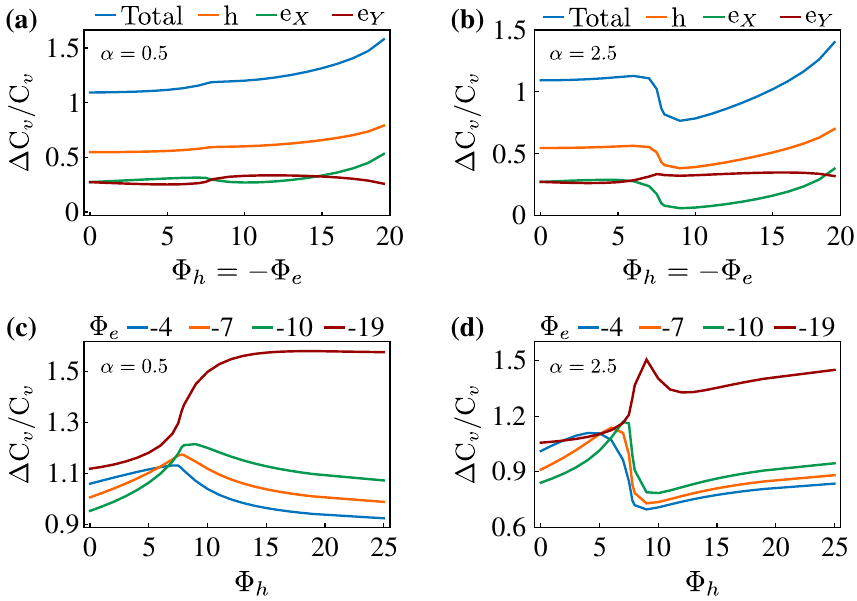}
 \caption{ The variation of the scaled specific heat jump $\frac{\Delta C_v}{C_{v}}$ with  hole nematic order $\Phi_h$ for (a,c) $\alpha=0.5$ and (b,d) $\alpha=2.5$. We set $\Phi_e=-\Phi_h$ in (a) and (b). In (c) and (d), we choose a set of values for $\Phi_e=\{-4, -7, -10,-19\}$ meV. }
\label{full cv with phi}
\end{figure}

The effect of nematicity on the specific heat jump is involved because $N_i$, $c_i$, and the coherence factors $\cos{\phi_i}$, all vary  with it. In Fig.~\ref{full cv with phi}, we plot $\Delta C_v/C_v$ as a function of $\Phi_h$ for various values of $\Phi_e$ and  representative $\alpha =0.5$ and $2.5$, chosen to be  smaller and larger than $\sqrt{2}$. For $\alpha =0.5$, we expect from Eq.~(\ref{final expression}) that $\Delta C_v/C_v \approx \Delta_1^4$, and we verified that the behavior of $\Delta C_v/C_v $  matches the behavior of $\Delta^4_1$ with $\Delta_1$ from Fig.~\ref{gap structure at small and large alpha}(a).

For $\alpha=2.5$, we expect $\Delta C_v/C_v \sim \Delta_2^4$, and the  behavior of $\Delta C_v/C_v$  matches the behavior of $\Delta^4_2$ with $\Delta_2$ from Fig.~\ref{gap structure at small and large alpha}(b). In both cases, we see that $\Delta C_v/C_v$ is generally around one, but increases with $\Phi_h$. Viewed as a function of  $\Phi_h$, $\Delta C_v/C_v$ displays a  kink like non-analyticity at $\Phi_h = \Phi_{cr}$ and, moreover, is non-monotonic at $\alpha =2.5$.  The non-monotonic behavior for this $\alpha$ is clearly visible in Fig.~\ref{full cv with phi}(d), where we plot $\Delta C_v/C_v$ vs. $\Phi_h$  for various $\Phi_e$. Fig.~\ref{full cv with phi}(c) shows that it also holds at $\alpha =0.5$, for large enough $|\Phi_e|$. At large $|\Phi_e|$ and even larger $\Phi_h$,  $\Delta C_v/C_v$ saturates.  The reason is that for such $\Phi$, the Y pocket mostly of $d_{xy}$ character and the X pocket is mostly of $d_{yz}$ character, hence $\langle \cos^a{\phi_Y}\rangle \ll 1$ and  $\langle \cos^a{\phi_X}\rangle \approx 1$, where $a =4,8$. Then $\Delta C_v/C_v \sim N_X/(N_h + N_X + N_X)$ and $N_X$ is the largest, see Fig.~\ref{DOS}(d). Note that for large $|\Phi_e| =19$ meV, $\Delta C_v/C_v$ is $1.5-1.6$.

For $\alpha \approx \sqrt{2}$, the behavior of $\Delta C_v/C_v$ vs $\Phi_h$  is intermediate between the
ones at $\alpha =0.5$ and $\alpha =2.5$.

 We also plot in Fig.~\ref{full cv with phi}(a,b) the band resolved contributions from hole and electron pockets.  We see that the largest contribution to the jump comes from the hole pocket. $\Delta C_v^h/C_v$  is non-analytic at $\Phi_{cr}$ and gives rise to non-analyticity in the full $\Delta C_v/C_v$.

\subsection{Specific heat jump at $T_c$ for scenario B \label{sec:scenarioB}}

\begin{figure}[t]
		\centering
		\includegraphics[width=1\linewidth]{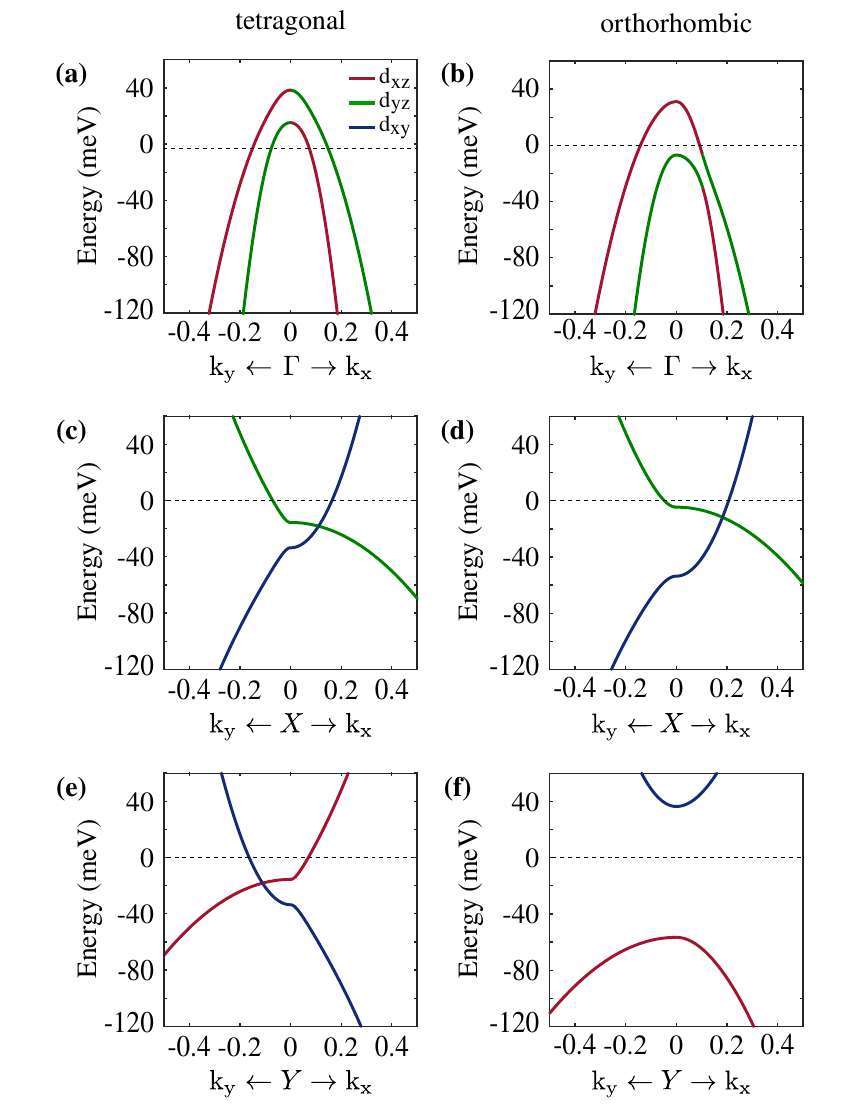}
		\caption{Scenario B: Calculated band dispersion of the 1-Fe unit cell in tetragonal and orthorhombic phase, respectively, near (a-b) $\Gamma$-, (c-d) $X$- and (e-f) $Y$- point. Fitting parameters taken from Ref.~[\onlinecite{rhodes2020non}]. Note that the $d_{xy}$ dominated Y-band is fully located above the Fermi level in (f).  }
		\label{Fig:1_1B}
\end{figure}

\begin{figure}[t]
	\centering
	\includegraphics[width=\columnwidth]{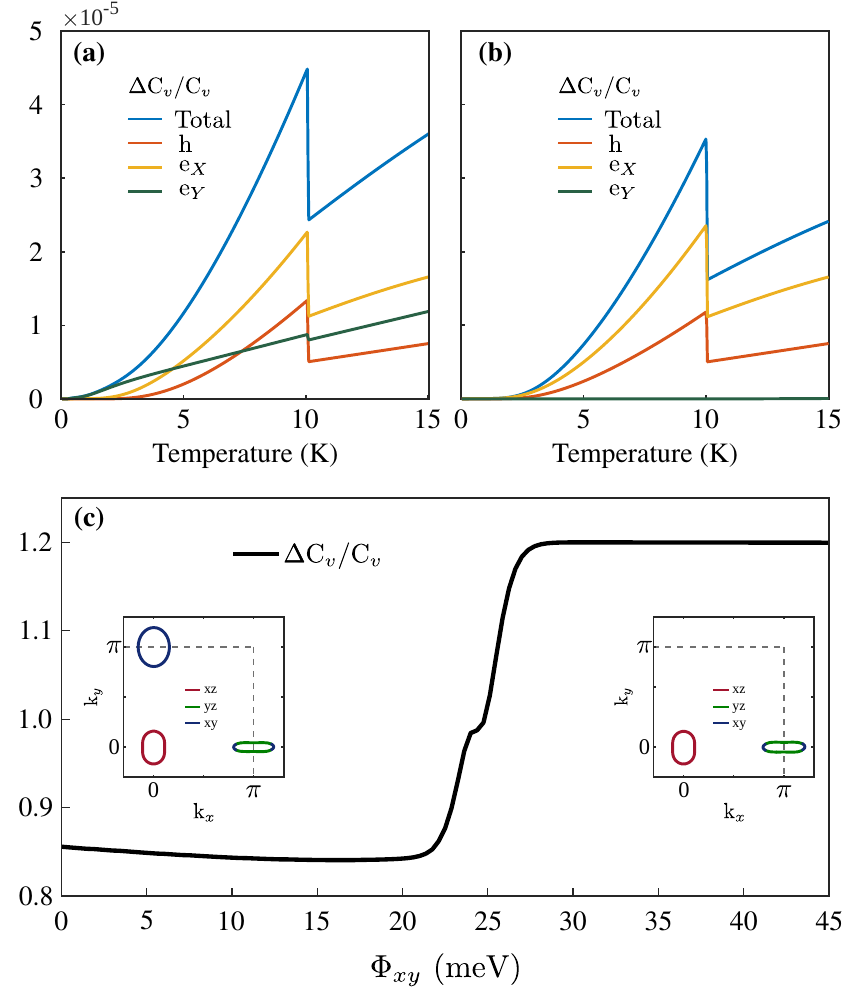}
	\caption{The total and band resolved specific heat calculated for (a) scenario A ($\Phi_{xy}=0$) and (b) scenario B ($\Phi_{xy}=45$ meV);  (c) $\Delta C_v/C_v$ as a function of $\Phi_{xy}$. For small (large)  $\Phi_{xy}$ scenario A (B) is valid. Left inset: Fermi surface for $\Phi_{xy} =0$. Right inset:  Fermi surface for $\Phi_{xy}=45$ meV. }
	\label{Fig:4}
\end{figure}

Below we present the results for the specific heat jump and its decomposition into contributions from different bands for the scenario B, when there is additional contribution $\Phi_{xy}$,  Eq.~(\ref{a_1_1}). This contribution splits the dispersions of $d_{xy}$ fermions on $X$ and $Y$ pocket. We choose the sign and magnitude of $\Phi_{xy}$ such that the bottom of the $Y$-band moves above the chemical potential, i.e., $Y$ pocket disappears in the nematic phase. To simplify calculations, we adopt the "antisymmetric approach" of Ref.~[\onlinecite{rhodes2020non}] and introduce $\Phi_{xy}$ nematic order only for $d_{xy}$ fermions on the Y pocket, as $2 \Psi_Y \Phi_{xy} \Psi_Y$ with $\Psi_Y$
from Eq.~(\ref{electro hamiltonian}).  Appropriate parameters to fit the band structure, available from ARPES experiments, in this scenario are given in the supplementary of Ref.~[\onlinecite{rhodes2020non}] and yield the Fermi surface shown in panel (d) of Fig.~\ref{Fig:0}(d) and in the right inset in Fig.~\ref{Fig:4}(c).
 The corresponding band dispersions at $\Gamma$-, $X$- and $Y$-points is shown in Fig.~\ref{Fig:1_1B}.

We solve the full non-linear gap equations (\ref{hole gap equation}-\ref{Y gap equations}), substitute the results into  Eq.~(\ref{Eq:Specificheat}) and obtain the specific heat.

In Fig.~\ref{Fig:4}(a)  we show the total specific heat $C_{V}$ (solid-blue) as well as the band resolved contributions from the $\Gamma$, $X$ and $Y$ pocket (solid yellow, green and orange, respectively). For definiteness we set $\Phi_{xy}=45$ meV, $\alpha =0.5$ and $J_{eh}=J_{ee}$ ($=U_{eh}/3$). We adjusted $U_{eh}$ to match experimental $T_c\sim10$ K. Observe that both $\Gamma$- and X-pocket contribute substantially to the specific heat jump, with the contribution from $Y$ pocket almost vanishes.  The largest contribution comes from the $X$ pocket.  This differs from the result for scenario A, but the difference is largely due to different parameters, as we verified.

A more substantial difference is actually  for the specific heat in the normal state. In scenario A $d_{xy}$ fermions from both $X$ and $Y$ pocket contribute to $C_v(T)$ above $T_c$. In scenario B, $d_{xy}$ fermions from $Y$ are gapped, and only $d_{xy}$ fermions from $X$ contribute.  As a result, the normal state $C_v(T)$ is reduced in scenario B compared to A, while $\Delta C_v$ at $T_c$ remains the same as only $d_{xz}$ and $d_{yz}$ fermions contribute to the jump. As the consequence, $\Delta C_v/C_v$ is larger in scenario B than in scenario $A$.  We show this explicitly where we plot $\Delta C_v/C_v$ as a function of  $\Phi_{xy}$ that drives the system between scenario A and scenario B.  We see that $\Delta C_v/C_v$ is roughly a constant at small $\Phi_{xy}$, when scenario A is valid. It then rapidly increases and saturates at a larger value at large $\Phi_{xy}$, when scenario B is valid.

\subsection{Comparison between scenario A, B and experiments \label{sec:comparison}}

Specific heat measurements in FeSe~[\onlinecite{chen2017highly, lin2011coexistence, hardy2019calorimetric, sun2017gap,Klein2019,Mizukami2020_arxiv,Karlsson_2015,Sun2018,Jiao2017,Rossler2018}] consistently reveal that $\Delta C_v/C_v \approx 1.65$. This is
 larger than the BCS result for a single band superconductor, $\Delta C_v/C_v\approx 1.43$. A larger $\Delta C_v/C_v$ is often associated with the effects beyond BCS~[\onlinecite{Marsiglio1986},\onlinecite{Carbotte1990}]. However, earlier works~[\onlinecite{Mishonov2005,Mishonov2005b,Nicol2005,Zehetmayer2003,Zehetmayer2013,Maiti_2010}] have found that in a multi-band system  $\Delta C_v/C_v$ can be either larger or smaller than the BCS value already within BCS approximation. In our analysis, we obtain $\Delta C_v/C_v$ around one in scenario A for small $\Phi_e$ and $\Phi_h$, but larger $\Delta C_v/C_v \sim 1.5-1.6$ for larger $\Phi_h \sim |\Phi_e| \leq 20$ meV. Within scenario B, $\Delta C_v/C_v$ is always larger than in the scenario A because the normal state contribution is smaller.  Then the experimental  $\Delta C_v/C_v \sim 1.65$ can be reproduced  already at smaller $\Phi_{h,e}$. In summary, the specific heat jump can be reproduced within both scenarios, but the parameter space is somewhat larger in scenario B.

\section{Specific heat near a possible transition into an $s+e^{i\eta} d$ state \label{sec:spec_heat_2}}

In this section we consider a possibility of a second superconducting transition in FeSe, caused by a transformation of the $s+d$ state into the $s+e^{i\eta}d$ state. Such an instability may arise near the point where the pairing interaction is attractive in both s-wave and d-wave channels, with comparable magnitudes.
The parameter range of $s+e^{i\eta}d$  has been previously analyzed in Ref.~[\onlinecite{kang2018time}], assuming that the nematic order is weak. Here we don't keep $\Phi$ small and include into consideration orbital transmutation in the nematic phase.  We  identify the parameter range, where $s+e^{i \eta} d$ order emerges.

\begin{figure}[t]
	\centering
	\includegraphics[width = \columnwidth]{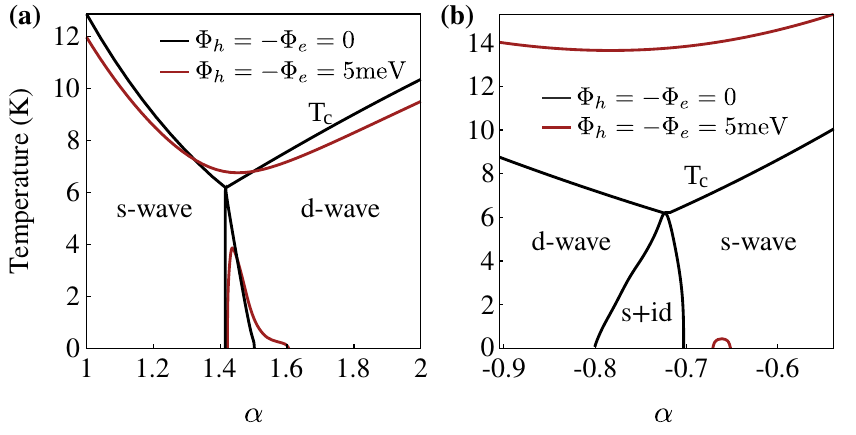}
	\caption{Regions of  the mixed $s+e^{i\eta}d$ order in the tetragonal and orthohombic phases  in the $(T, \alpha)$ plane.  In (a) we set $J_{ee}=0$ and $J_{eh}<0$ to bring $\alpha = (U_{eh} - J_{eh})/(U_{eh} + J_{eh})$ close to $\sqrt{2}$. In (b) we set $J_{ee}=J_{eh} >0$. A $d$-wave order develops when $J_{eh}/U_{eh}$ is larger than a certain number. For our parameter, the mixed phase is located near $\alpha \sim -0.7$. The shrinking of the range of $s+e^{i\eta}d$ order with nematicity is stronger in (b) than in (a).}
	\label{Fig:2}	
\end{figure}

\begin{figure}[t]
	\centering
	\includegraphics[width = \columnwidth]{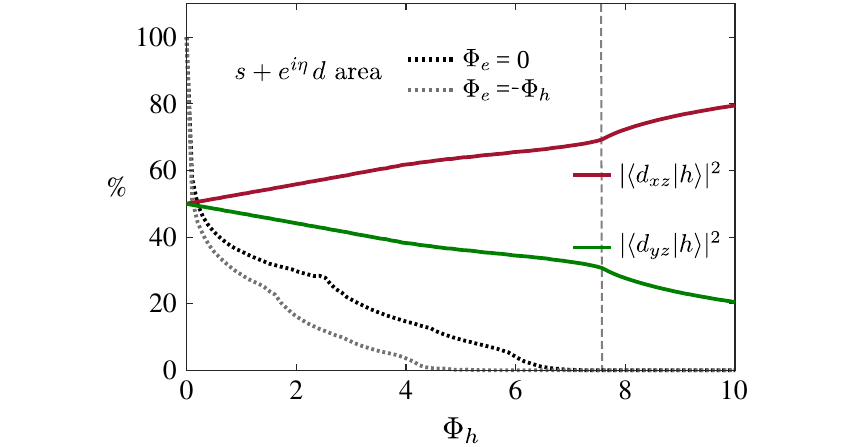}
	\caption{ The size of $s+e^{i \eta}d$ region (black) and the percentage of $d_{xz}$ and $d_{yz}$  orbital content at the hole pocket (red and green), vs $\Phi_h$.
Black-dashed line is for $\Phi_e=0$,  gray-dashed one is for $\Phi_e=-\Phi_h$.}
	\label{Fig:3}
\end{figure}

To analyze the transition to  $s+e^{i\eta}d$-wave state, we numerically solve the full  non-linear gap equations (\ref{hole gap equation}-\ref{Y gap equations}), including both $s$-wave and $d$-wave harmonics. We show our results in Fig.~\ref{Fig:2}.
In panel (a), we assume $J_{ee}=0$ and vary the parameter $\alpha$, which drives the system from s- to d-wave symmetry at $\alpha \approx \sqrt{2}$. In panel (b) we set $J_{ee}=J_{eh}$, in this case the transition from s- to d-wave is at negative $\alpha \sim -0.7$. The black curves in the Fig.~\ref{Fig:2} are the results for $\Phi=0$. In both panels, there is a sizable range of $s+id$ order, sandwiched between pure $s$-wave and $d$-wave states.  This is consistent with Ref.~[\onlinecite{kang2018time}]. For a finite nematic order, the gap function in the mixed state is $s+ e^{i\eta}d$, where $0\leq\eta\leq \frac{\pi}{2}$. The results for $\Phi \neq 0$ show that nematicity generally suppresses the width of the $s+ e^{i\eta}d$ region, but the suppression is far stronger for $J_{ee} = J_{eh}$ (panel b) than for $J_{ee} =0$ (panel a). The reason why a nematic order is unfavorable for the $s+ e^{i\eta} d$ state is again orbital transmutation: as we said a nematic order makes pockets ”mono-orbital” and therefore favors $s$-wave pairing. Consequently, the region, where s- and d-wave pairing channels are nearly degenerate, gets suppressed. We illustrate this in Fig.~\ref{Fig:3}, where we plot the area of $s+e^{i \eta}d$ region, normalized to its value in the tetragonal state, and the difference in the orbital content on the hole pocket, both vs $\Phi_h$.
 We see that the area of the mixed range shrinks and vanishes when $\Phi_h$ reaches $\Phi_{cr}$.

Specific heat measurements on FeSe in Ref.~[\onlinecite{chen2017highly,sun2017gap,Sun2018,Jiao2017}] reported two jumps at $T_c=8$K and at $T^*\sim1$K. The jump at $T_c$ clearly indicates the transition to the superconducting phase. In Ref.~[\onlinecite{kang2018time}] it was argued that the jump at $T^*=1$K  might be explained by the  transition into the $s+\text{e}^{i\eta}d$ phase. Our results show that this is possible, but unlikely as the parameter range when $s+\text{e}^{i\eta}d$ order develops is quite narrow.

We also note in passing that in panel (b) of Fig.~\ref{Fig:2},  $T_c$ goes up at a nonzero $\Phi$, despite that a nematic order is generally believed to be a competitor to superconductivity. This happens because  $J_{eh}$ is the dominant component of the pairing interaction, and $J_{eh}$ couples  $d_{xz}$  fermions on the $\Gamma$-pocket to  $d_{yz}$ fermions on the $X$-pocket. The spectral weight of  both fermions get enhanced by sign-changing $d_{xz/yz}$ nematicity, and this enhances $T_c$.  For the case in panel (a), the dominant interaction is $U_{eh}$ that couples $d_{xz}$ ($d_{yz}$) orbitals at $\Gamma$ with $d_{xz}$ ($d_{yz}$) orbitals at $Y$ ($X$). In the nematic phase $d_{xz}$ ($d_{yz}$) weight is enhanced (reduced) at $\Gamma$ but reduced (enhanced) at $Y$ ($X$). As a consequence, $T_c$ is weakly affected by nematicity.

\section{Conclusions}

\label{sec:conclusion}
In this paper we presented in-depth analysis of superconducting gap function and specific heat of a multi-orbital metal, like FeSe, which first develops a nematic order and then undergoes a transition into a superconducting state, which co-exists with nematicity.  We considered two scenarios:  scenario A, in which nematic order develops between $d_{xz}$ and $d_{yz}$ orbitals on hole and electron pockets ($\Phi_h$ and $\Phi_e$)  and scenario B, in which there is an additional component of the nematic order for $d_{xy}$ fermions on the two electron pockets ($\Phi_{xy}$).

We specifically addressed three questions. The first one is the angular dependence of the gap. Here we analyzed the competition between the two effects. One is nematicity-induced $s-d$ mixture, which necessary induces angular variation of the gap function even if the superconducting state is an s-wave without nematicity. Another one is orbital transmutation of low-energy excitations in the nematic state. This effect tends to make Fermi surface pockets  mono-orbital and thus favors an angle-independent gap function. We analyzed the crossover from initial $s-d$ mixing to eventual angle-independent gap and argued that the most likely scenario for stronger $s$-wave attraction in the tetragonal phase is a gap function with no nodes, while for stronger $d$-wave attraction the 4 nodes from $d$-wave order disappear once nematic order exceeds a certain threshold.  However, in a  parameter range where  $s$-wave and $d$-wave interactions have comparable strength, we find more involved crossovers in which, e.g., the number of nodal points goes from zero to a finite number and then back to zero, or when the number of nodal points goes from $4$ to $8$ and then to zero.

The second question that we addressed is the behavior of a specific heat in a nematic superconductor. For this, we solved the non-linear gap equation, obtained the forms of the gaps below $T_c$, and used them to compute  the specific heat $C_v(T)$. We analyzed the evolution  of $C_v(T)$ with the nematic order in both the scenario A and the scenario B.  Here our key result is the specific heat jump at $T_c$: $\Delta C_v/C_v$. We found that $\Delta C_v/C_v$ is around one in the tetragonal phase, for parameters appropriate for FeSe. The magnitude of $\Delta C_v/C_v$  increases with the nematic order and saturates at $\Delta C_v/C_v \sim 1.5-1.6$. This is quite consistent with the experimental result for FeSe $\Delta C_v/C_v \sim 1.65$  (Refs.~[\onlinecite{chen2017highly, lin2011coexistence, hardy2019calorimetric, sun2017gap,Klein2019,Mizukami2020_arxiv,Karlsson_2015,Sun2018,Jiao2017,Rossler2018}]).  The values of $\Phi_{h,e}$  requires to  reach saturation are smaller in scenario B as in this scenario the normal state $C_v(T)$ is smaller as it assumes that the Y pocket  disappears because of sizable $\Phi_{xy}$.

The third question that we addressed is a potential transition at $T_{c1} < T_c$ from an $s+d$ state to an $s + e^{i\eta} d$ state that breaks time-reversal symmetry.  Such a transition was  suggested~[\onlinecite{kang2018time}] as a possible explanation of the experiments~[\onlinecite{chen2017highly,sun2017gap,Sun2018,Jiao2017,Klein2019}], which observed a second jump of  $C_v(T)$ at $T^* \sim 1K$, well below $T_c \sim 8.5 K$.  At small $\Phi_{h,e}$, previous study found~[\onlinecite{kang2018time}] that the parameter range where $s + e^{i\eta} d$ state  develops at $T \to 0$ is quite sizable.  We analyzed larger $\Phi_{h,e}$ and found that the range shrinks due to orbital transmutation which acts against competition between $s$- and $d$-pairing. We expect that the  measurements of the gap function and specific heat in doped FeSe$_{1-x}$S$_x$ or FeSe$_{1-x}$Te$_x$, where the amount of nematic order varies with $x$, could verify the presence of the $s + e^{i\eta} d$ state. \newline

\section{Acknowledgments}
We thank Rafael Fernandes, Thierry Klein,  and Hai-Hu Wen for useful conversations. The work by KRI and AVC was supported  by US Department of Energy, Office of Science, Basic Energy Sciences, under Award
No. DE-SC0014402.  The work of JB and IME was supported by the joint NSFC-DFG grant (ER 463/14-1)\\

$^*$ These two authors equally contributed to this work.

\bibliography{ref}	
\appendix
\begin{widetext}
\section{Singularities in the hole coherence factor}
\label{singularities}
In this section, we compute $\langle \cos{2 \phi_h}\rangle$ and $\langle \cos^2{2 \phi_h}\rangle$ as a function of $\Phi_h$ and show, respectively, that they exhibit an $x\ln{|x|}$ and $x^2\ln{|x|}$ type non-analyticity near the critical nematic strength $\Phi_{cr}$(defined below). Using Eqs.(\ref{hole disperison},\ref{cos2phih}), we write $\cos{2 \phi_h}$ on the Fermi surface as,
\begin{equation}
\cos{2 \phi_h}=\frac{b \frac{\kk_F(\theta)^2}{2}\cos(2\theta_h)-\Phi_h}{ \frac{\kk_F(\theta)^2}{2 m_h}-\mu_h}
\label{cos2phihonFS}
\end{equation}
Here, $\kk_F(\theta)$ is the Fermi radius at an angle $\theta$. We define $\frac{\kk_F^2}{2}=x_f(\theta)$ for convenience, and write
\begin{align}
\cos{2 \phi_h}&=\frac{b x_f(\theta)\cos(2\theta_h)-\Phi_h}{\frac{x_f(\theta)}{ m_h}-\mu_h}\nn\\
&= m_h b \frac{b x_f(\theta)\cos(2\theta_h)-\Phi_h}{b  x_f(\theta) -\Phi_{cr}},
\label{cos2phihonFS}
\end{align}
where $\Phi_{cr}= \mu_h m_h b$ is the critical nematic strength where the orbital order  in $k_x$-direction changes from $d_{yz}$ to $d_{xz}$ on the hole pocket.  We set $m_h b=t$ for convenience and for our model parameters from TABLE-\ref{table:Table S1}, $t \approx 0.5$. We find the functional form of $b x_f(\theta)$  from  the band dispersion Eq.~(\ref{hole disperison}) as,
\begin{equation}
b x_f(\theta)=\frac{\Phi_{cr}-t^2 \Phi_h \cos{2\theta}+\sqrt{\left(\Phi_{cr}-t^2 \Phi_h \cos{2\theta} \right)^2-\left(1-t^2\right)\left(\Phi_{cr}^2-\Phi_h^2 t^2\right)}}{1-t^2}.
\label{bf}
\end{equation}

In the limit where nematic order is small, i.e. $\Phi_h\ll\Phi_{cr}$ we can expand Eq.~(\ref{bf}) and Eq.~(\ref{cos2phihonFS}) in $\frac{\Phi_h}{\Phi_{cr}}\ll1$, which yields
\begin{eqnarray}
\langle \cos{2\phi_h}\rangle_{FS} &=& -\frac{1-t}{2} \left(\frac{\Phi_h}{\Phi_{cr}}\right)-\frac{1-t^2}{16} \left(\frac{\Phi_h}{\Phi_{cr}}\right)^3+O(\Phi_h^5),\\
\langle \cos^2{2\phi_h}\rangle_{FS} &=& \frac{1}{2}-t \frac{1-t}{4} \left(\frac{\Phi_h}{\Phi_{cr}}\right)^2+O(\Phi_h^4).
\end{eqnarray}
\newline
 In the limit when $\Phi_h \approx \Phi_{cr}$, we find from Eq.~(\ref{bf}) precisely at $\Phi_h= \Phi_{cr}$
\begin{equation}
b x_f^{cr}(\theta,\Phi_{cr})=\Phi_{cr}+\Phi_{cr} f(\theta),
\end{equation}
where
\begin{equation}
f(\theta)=\frac{2 t^2 \sin(\theta)^2+2 t |\sin(\theta)|\sqrt{1-t^2 \cos(\theta)^2}}{1-t^2}.
\end{equation}
Then,
\begin{align}
\cos{2 \phi_h}&=t \frac{\left(\Phi_{cr}+\Phi_{cr} f(\theta)\right) \cos(2\theta)-\Phi_{cr}}{\Phi_{cr}+\Phi_{cr} f(\theta)-\Phi_{cr}}\nn\\
&= t \left[-\frac{1}{f(\theta)}+\cos(2\theta)\left(1+\frac{1}{f(\theta)}\right)\right]\nn\\
&= t \left[\cos(2\theta) -2 \frac{\sin(\theta)^2}{f(\theta)}\right].
\label{co2phihatphicr}
\end{align}
 From Eq.~(\ref{co2phihatphicr}), we find that near $k_x$-axis, $\cos{2\phi_h}$ approaches the value $t$, while it  is undefined  in $k_x$-direction. We will show later $\cos{2\phi_h}(0)=\text{sgn}(\Phi_{cr}-\Phi_{h})$
\begin{equation}
\lim_{\theta\rightarrow 0} \cos{2\phi_h}(\theta)_ {\Phi_{cr}}=t.
\end{equation}
Averaging $\cos{2\phi_h}$ and $\cos^2{2\phi_h}$ over the angle $\theta$, we get,
\begin{eqnarray}
\langle \cos{2\phi_h}\rangle &=& t \langle\cos(2\theta)\rangle -2t \left \langle\frac{\sin(\theta)^2}{f(\theta)} \right\rangle\nn\\
&=& \frac{t}{2}-\frac{\sqrt{1-t^2}}{\pi}-\frac{\arcsin(t)}{\pi t}\nn\\
&=& \approx -0.32,
\label{zero point value}
\end{eqnarray}

\begin{eqnarray}
\langle \cos^2{2\phi_h}\rangle &=& t^2 \left\langle\cos^2{2\theta} +4\dfrac{\sin^4{\theta}}{f(\theta)^2} -4 \dfrac{\cos{2\theta} \sin^2{\theta}}{f(\theta)}\right\rangle\nn\\
&=&  \dfrac{t(2(2-2t^2)\sqrt{1-t^2}+\pi t(2+t^2))-2 \arcsin{t}}{4 \pi t^2}\nn\\
&=& \approx 0.36.
\label{zero point value for cossq}
\end{eqnarray}

Next we assume  $\Phi_h=\Phi_{cr}+\delta$ and show  how $\langle \cos{2 \phi_h}\rangle$ and $\langle \cos^2{2 \phi_h}\rangle$ depend on $\delta$.
Using Eq.~(\ref{bf}), we show,
\begin{equation}
b x_f(\theta,\delta)=\frac{\Phi_{cr}\left(1-t^2 \cos(2\theta)\right)-\delta t^2 \cos(2\theta)+\sqrt{B}}{1-t^2},
\label{delta bxf}
\end{equation}
where, 
\begin{equation}
B=4 \Phi_{cr}^2 t^2(1-t^2 \cos(\theta)^2)\sin(\theta)^2+\delta^2 t^2- 4 \delta^2 t^4 \sin(\theta)^2 \cos(\theta)^2+4 \delta t^2 \Phi_{cr}\sin(\theta)^2(1-2 t^2 \cos(\theta)^2).
\end{equation}
At $\theta= 0, \pi$, $b x_f(0,\delta)$ has a $\left|\delta\right|$ type non-analyticity  as we find from Eq.~(\ref{delta bxf})
\begin{equation}
b x_f(0, \delta)=\Phi_{cr}+\frac{t |\delta|}{1+t \hspace{.1cm} \text{sgn}(\delta)}
\end{equation}
 and plot in Fig.~\ref{Fig.17}.
\begin{figure}[h]
\centering
\includegraphics{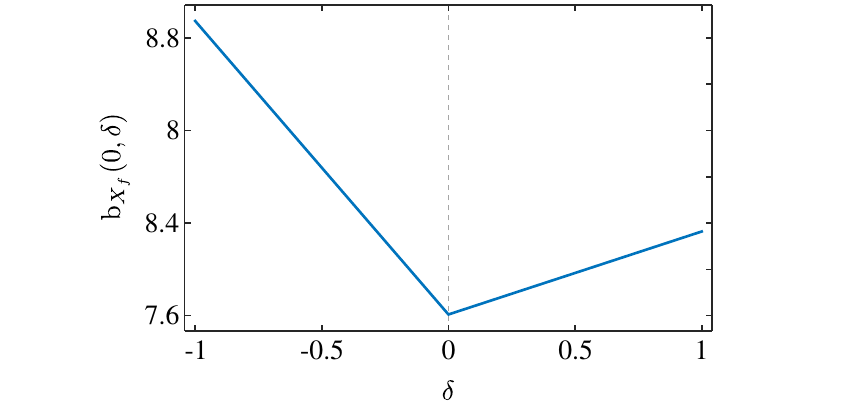}
\caption{$b x_f(0, \delta)$ as function of $\delta$.}
\label{Fig.17}
\end{figure}
As a result, we find,
\begin{eqnarray}
\cos{2\phi_h}\left(0\right) &= &t\dfrac{b x_{f}(0,\delta)-\Phi_{cr}-\delta}{b x_f(0,\delta)-\Phi_{cr}}\nn\\
&=& t \left(1-\dfrac{\delta}{b x_{f}(0,\delta)-\Phi_{cr}}\right)\nn\\
&=& t \left(1-\dfrac{\delta(1-t^2)}{|\delta| t-\delta t^2}\right)\nn\\
&=& -\text{sgn}(\delta)
\end{eqnarray}
Next, we move to calculate $\langle \cos{2 \phi_h}\rangle$,
\begin{eqnarray}
\langle \cos{2 \phi_h}\rangle &=& t \left\langle \frac{b x_f(\theta,\delta)\cos(2\theta)-\Phi_h}{b x_f(\theta,\delta)-\Phi_{cr}}\right\rangle\nn\\
&=& t \left\langle \frac{b x_f(\theta,\delta) \cos(2\theta)-\Phi_{cr}-\delta}{b x_f(\theta,\delta) -\Phi_{cr}}\right\rangle\nn\\
&=& t \underbrace{\left\langle \frac{b x_f(\theta,\delta) \cos(2\theta)-\Phi_{cr}}{b x_f(\theta,\delta) -\Phi_{cr}}\right\rangle}_{\text{Term 1}} -\delta t \underbrace{\left\langle \frac{1}{b x_f(\theta,\delta) -\Phi_{cr}}\right\rangle}_{\text{Term 2}}.
\label{delta eq}
\end{eqnarray}

We claim that the second blue under-braced term in Eq.~(\ref{delta eq}) contains the non-analytic  behavior of of $\langle \cos{2\phi_h}\rangle$, because, as $\delta\rightarrow 0$, the denominator diverges at $\theta=0$ and $\pi$, see Fig.~\ref{terms}(b). The first under-braced term in Eq.~(\ref{delta eq}) is almost independent of variations in $\delta$, as can be seen in Fig.~\ref{terms}(a).
\begin{figure}[h]
\includegraphics{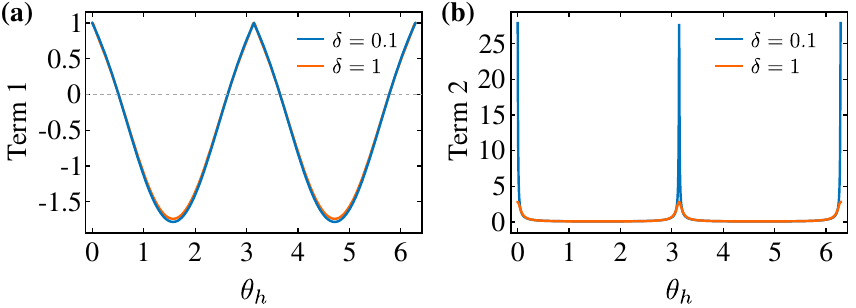}
   \caption{We plot both of the integrands of Eq.~\ref{delta eq} as a function of $\theta$ for $\delta=0.1$ and $1$. }
    \label{terms}
  \end{figure}
Hence, we approximate the first term of Eq.~(\ref{delta eq}) by setting $\delta=0$ and  recover result of Eq.~(\ref{zero point value}).\newline

To calculate the second term, we rewrite the denominator to separate the singular from the regular part,
  \begin{eqnarray}
  b x_f(\theta,\delta)-\Phi_{cr}&=&b x_f(\theta,\delta)-b x_f(0,\delta)+b x_f(0,\delta)-\Phi_{cr}\nn\\
  &=& \frac{t |\delta|}{1+ t \hspace{.1cm} \text{sgn}(\delta)}+\left( b x_f(\theta,\delta)-b x_f(0,\delta)\right).
  \label{Seperation}
  \end{eqnarray}
We again approximate that $\left( b x_f(\theta,\delta)-b x_f(0,\delta)\right)$ does not change much with $\delta$. So we write it as,
\begin{equation}
   b x_f(\theta,\delta)-b x_f(0,\delta)  \approx b x_f(\theta,0)-b x_f(0,0)=\phi_{cr} f(\theta).
\end{equation}
Then, the  non-analytic contribution of $\langle \cos{2\phi_h}\rangle$ is,
\begin{eqnarray}
 \langle \cos{2\phi_h}\rangle &=& -\delta  \left\langle\frac{1}{\Phi_{cr} f(\theta)+\frac{t |\delta|}{1+ t \hspace{.1cm} \text{sgn}(\delta)}}\right\rangle\nn\\
 &=& -x  \left\langle\frac{1}{  f(\theta)+ A(x)}\right\rangle\label{To Integrate},
\end{eqnarray}
where, $x=\frac{\delta}{\Phi_{cr}}$, and
\begin{equation}
  A(x)=t \frac{ |x|}{1+t \hspace{.1cm} \text{sgn}(x)}.
\end{equation}
We peform the integration over $\theta$ in Eq.~\ref{To Integrate}, and obtain
  \begin{align}
 \langle \cos{2\phi_h}\rangle = -2 x\left[\underbrace{\frac{\arccos(t)}{A(x)-1}}_{\textbf{Term 1}}+\underbrace{\frac{2-A(x)}{1-A(x)}\frac{\sqrt{1-t^2}}{\sqrt{(2-A(x))^2 t^2-A(x)^2}} \arctanh\left(\frac{\sqrt{(2-A(x))^2 t^2-A(x)^2}}{(2-A(x))t}\right)}_{\textbf{Term 2}} \right].
 \label{singularity}
 \end{align}
 As $\delta \rightarrow 0$, the first term of Eq.~(\ref{singularity}) inside the parenthesis  approaches a finite value($-\arccos(t))$, while the second term blows up because of the $\arctanh$ function(see Fig.~\ref{ap51}). We neglect the regular part, and expand the second term around $x=0$ to find the nonanalytic component which is of  $ |x| \log(|x|)$ form.
 \begin{figure}[h]
  \centering
  \includegraphics{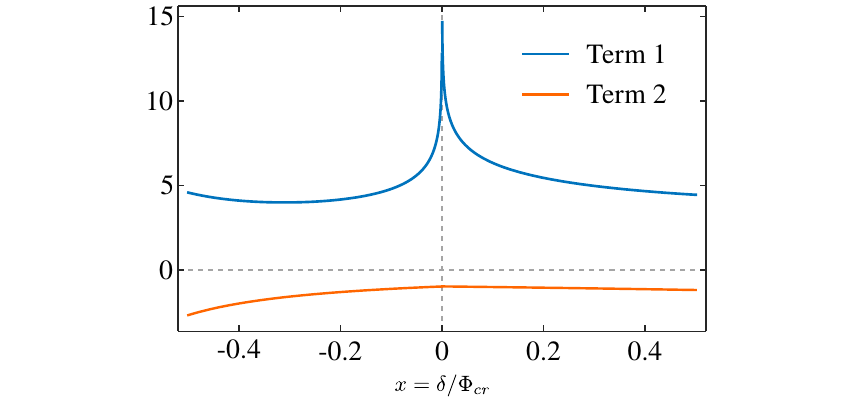}
  \caption{  Term 1 and Term 2 according to Eq.~\ref{singularity} as function of $x=\delta/\Phi_{cr}$}
  \label{ap51}
 \end{figure}
 \begin{equation}
   \langle \cos{2\phi_h}\rangle \propto -2 x\left(2+t \frac{|x|}{1+t \hspace{.1cm} \text{sgn}(x)}\right)\left(\frac{1}{2t}+\frac{|x|}{4(1+t\hspace{.1cm} \text{sgn}(x))}\right) \left(c-\log(|x|)\right) \propto |x| \log(|x|)
 \end{equation}
 Next, we compute $\langle \cos^2{2\phi_h}\rangle$ in the following way,
\begin{eqnarray}
  \langle \cos^2{2\phi_h}\rangle &=& t^2 \left\langle \frac{b x_f(\theta,\delta)\cos(2\theta)-\Phi_h}{b x_f(\theta,\delta)-\Phi_{cr}}\right\rangle^2\nn\\
&=& t^2 \left[ \left\langle \frac{\left(b x_f(\theta,\delta) \cos(2\theta)-\Phi_{cr}\right)^2}{\left(b x_f(\theta,\delta) -\Phi_{cr}\right)^2}\right\rangle +\delta^2 \left\langle \dfrac{1}{\left(b x_f(\theta,\delta) -\Phi_{cr}\right)^2}\right\rangle \right.  \left. -2 \delta \left\langle \frac{\left(b x_f(\theta,\delta) \cos(2\theta)-\Phi_{cr}\right)}{\left(b x_f(\theta,\delta) -\Phi_{cr}\right)^2}\right\rangle \right]\nn\\
\label{delta2 eq}
\end{eqnarray}

The first term of Eq.~(\ref{delta2 eq}) contains no singularity and gives the $\delta=0$ contribution to $\langle\cos^2{2\phi_h}\rangle$. To calculate the singularity present in the second term, we approximate the denominator as we did in Eq.~(\ref{Seperation}).  We further approximate the function $f(\theta)$ near $\theta=0$, where the non-analiticity is located and find
\begin{equation}
   f(\theta)=\dfrac{2 t}{\sqrt{1-t^2}}\left(\theta+ A_2 \theta^2+O(\theta^3)\right),
   \label{approximate ff}
\end{equation}
where, $A_2=\dfrac{t}{\sqrt{1-t^2}}$. Using Eq.~(\ref{approximate ff}), we calculate the second term of Eq.~(\ref{delta2 eq}) as
\begin{eqnarray}
   t^2 \delta^2 \left\langle \dfrac{1}{\left(b x_f(\theta,\delta) -\Phi_{cr}\right)^2}\right\rangle &=&(1-t^2) x^2 \int_0^{\pi/2}\dfrac{1}{\left(\theta+ A_2 \theta^2+ A_0(x)\right)^2}.
   \label{second term int}
\end{eqnarray}
We define $A_0(x)=\dfrac{\sqrt{1-t^2}}{2} A(x)$.  Eq.~(\ref{second term int}) can be computed exactly, and is equal to,
\begin{eqnarray}
   x^2\left(\dfrac{\pi}{A_0(x)\left(4 A_0(x)+\pi \left( 2+A_2 \pi \right)\right)}-\dfrac{\pi\left(1+A_2\pi\right)}{A_0(x) V(x)^2 \left(4 A_0(x)+\pi \left( 2+A_2 \pi \right)\right)} \right. \nn\\ \left. -4 \dfrac{A_2}{V(x)^3} \left[\arctan\left(\dfrac{1}{V(x)}\right)-\arctan\left(\dfrac{1+A_2\pi}{V(x)}\right)\right]\right).
   \label{delta sq term}
\end{eqnarray}
We define $V(x)=\sqrt{-1+4 A_2 A_0(x)}$. When $x\rightarrow 0$, $V(x)\rightarrow i$. The first and second term of Eq.~(\ref{delta sq term}) are regular. To identify the non-analytic behaviour of the third term, we use the following identity Eq.~(\ref{arc tan}), and expand $V(x)$ upto the linear order in x,
\begin{align}
   \arctan(z)&=-\dfrac{i}{2}\log\left( \dfrac{1+i z}{1-iz}\right),   \label{arc tan}\\
   V(x)&=i\left(1-t^2 \dfrac{|x|}{1+ t \hspace{.1cm} \text{sgn}(x)}\right)=i \Tilde{V}(x),
   \label{v(x)}
\end{align}
where $\Tilde{V}(x)=1-t^2 \dfrac{|x|}{1+ t \text{sgn}(x)}$. Using Eq.~(\ref{arc tan},\ref{v(x)}), we find that,
\begin{equation}
   \arctan\left(\dfrac{1}{V(x)}\right)=-\dfrac{i}{2}\log\left(\dfrac{\Tilde{V}(x)+1}{\Tilde{V}(x)-1}\right)\propto \log\left(\dfrac{t^2|x|}{1+t \hspace{.1cm}\text{sgn}(x)}\right).
   \label{arctanv}
\end{equation}
Eq.~(\ref{arctanv}) shows that the most singular correction of Eq.~(\ref{delta sq term}) is of the form $x^2 \log(|x|)$. \newline \newline
Finally, we write the last term of the Eq.~(\ref{delta2 eq}) in the following way to show that it is also singular of the form $x \log(x)$.

\begin{eqnarray}
   \delta \left\langle \frac{\left(b x_f(\theta,\delta) \cos(2\theta)-\Phi_{cr}\right)}{\left(b x_f(\theta,\delta) -\Phi_{cr}\right)^2}\right\rangle
    &=& \delta \left\langle \frac{\left(b x_f(\theta,\delta)\left(1-2\sin^2{\theta}\right)-\Phi_{cr}\right)}{\left(b x_f(\theta,\delta) -\Phi_{cr}\right)^2}\right\rangle\nn\\
   &=& \delta \left\langle \frac{1}{\left(b x_f(\theta,\delta) -\Phi_{cr}\right)}\right\rangle- 2 \delta \left\langle \frac{\left(b x_f(\theta,\delta) \sin^2{\theta}\right)}{\left(b x_f(\theta,\delta) -\Phi_{cr}\right)^2}\right\rangle
\end{eqnarray}
We show that the first term  is singular of the form $x \log(|x|)$. We assume that the second term is not singular because of the $\sin^2{\theta}$ term in the numerator.

\section{BCS-gap equations\label{Appendix:GapEquations}}

We treat Eq.~(\ref{Pairing Hamiltonian}) in mean field approximation and obtain the BCS-gap equations for the band-space gaps as
\begin{align}
-\Delta_h(\kk)&= \left(U_s+U_d \cos{2\phi_h(\kk)}\right)\int_{\pp} \frac{\tanh{ \frac{E_X(\pp)}{2T}}}{2 E_X(\pp)}\cos^2{\phi_X(\pp)} \Delta_X(\pp)\nn\\
&+ \left(U_s-U_d \cos{2\phi_h(\kk)}\right)\int_{\pp} \frac{\tanh{ \frac{E_Y(\pp)}{2T}}}{2 E_Y(\pp)}\cos^2{\phi_Y(\pp)}\Delta_Y(\pp)\label{hole gap equation}\\
-\Delta_X(\kk)&=\cos^2{\phi_X(\kk)} \left[\int_{\pp} \frac{\tanh{\frac{E_h(\pp)}{2T}}}{2 E_h(\pp)}\left(U_s+U_d\cos{2 \phi_h(\pp)}\right) \Delta_h(\pp)+J_{ee}\int_{\pp}  \frac{\tanh{ \frac{E_Y(\pp)}{2T}}}{2 E_Y(\pp)}\cos^2{\phi_Y(\pp)} \Delta_Y(\pp)\right]\label{X gap equation}\\
-\Delta_Y(\kk)&=\cos^2{\phi_Y(\kk)}\left[\int_{\pp} \frac{\tanh{ \frac{E_h(\pp)}{2T}}}{2 E_h(\pp)}\left(U_s-U_d\cos{2 \phi_h(\pp)}\right) \Delta_h(\pp)+J_{ee} \int_{\pp} \frac{\tanh{\frac{E_X(\pp)}{2T}}}{2 E_X(\pp)}\cos^2{\phi_X(\pp)} \Delta_X(\pp)\right],
\label{Y gap equations}
\end{align}
where $E_i(\textbf{p})=\left(\xi_i^2(\textbf{p})+|\Delta_i((\textbf{p}))|^2\right)^{1/2}$ is the typical Bogoliubov quasiparticle spectrum and momentum integration is confined to an energy interval $\left[-\Lambda,\Lambda \right]$ around the Fermi surface.
Near $T_c$ the linearized gap equations are
\begin{align}
\Delta_1+\Delta_2 \cos{2\phi_h}&=- \log \frac{\Lambda}{T_c}\left[N_X \Delta_3 \left(U_s+U_d \cos{2\phi_h}\right) \langle \cos^4{\phi_X}\rangle  + N_Y \Delta_4\left(U_s-U_d \cos{2\phi_h}\right)  \langle \cos^4{\phi_Y}\rangle\right]\label{hole equation}\\
\Delta_3 &= - \log\frac{\Lambda}{T_c}\left[N_h \langle \left(U_s+U_d \cos{2\phi_h}\right)\left(\Delta_1+\Delta_2 \cos{2\phi_h}\right)\rangle + N_Y  \Delta_4 J_{ee}\langle \cos^4{\phi_Y}\rangle \right]\label{X equation}\\
\Delta_4 &= - \log\frac{\Lambda}{T_c}\left[ N_h \langle \left(U_s-U_d \cos{2\phi_h}\right)\left(\Delta_1+\Delta_2 \cos{2\phi_h}\right)\rangle + N_X \Delta_3 J_{ee}\langle \cos^4{\phi_X}\rangle \right]\label{Y equation}
\end{align}

\section{ Dependence of $\frac{\Delta_2}{\Delta_1}$ on the nematic order}
\label{delta2 delta1 ratio}

We set $J_{ee}=0$ in this section, and compute the ratio $\frac{\Delta_2}{\Delta_1}$ analytically. The largest eigenvalue $\lambda$ of the matrix equation(\ref{linearized gap equation full}) corresponding to the leading superconducting instability turns out to be,
\begin{align}
\lambda =
 \Bigg[\frac{N_h}{2} \Big[& g_0+2 \alpha \langle \cos{2\phi_h}\rangle g_1+\alpha^2 \langle \cos^2{2\phi_h}\rangle g_0+\nn\\
 &+ \sqrt{4 \alpha^2 \left(\langle \cos{2\phi_h}\rangle^2-\langle \cos^2{2\phi_h}\rangle\right)\left(g_0^2-g_1^2\right) +\left(g_0+2 \alpha \langle \cos{2\phi_h}\rangle g_1+\alpha^2 \langle \cos^2{2\phi_h}\rangle g_0\right)^2} \Big]\Bigg]^{1/2},
\end{align}
where
\begin{eqnarray}
g_0=N_X \langle \cos^4{\phi_X}\rangle + N_Y \langle \cos^4{\phi_Y}\rangle
\end{eqnarray}
and
\begin{eqnarray}
g_1=N_X \langle \cos^4{\phi_X}\rangle- N_Y \langle \cos^4{\phi_Y}\rangle
\end{eqnarray}
In the tetragonal phase, $g_1=0$ and $g_0= 2 N_X \langle \cos^4{\phi_X}\rangle$ (for our band parameters, $g_0\approx 0.1$). With increasing electron nematic order $\Phi_e$, $\langle \cos^4{\phi_Y}\rangle$ decreases since Y-pocket becomes mostly of $d_{xy}$ nature. As a result, $g_0-g_1$ decreases with $\Phi_e$.\\
To calculate the ratio $\frac{\Delta_2}{\Delta_1}$, we rewrite Eq.~(\ref{hole equation}),
\begin{eqnarray}
\Delta_2 &=& -\dfrac{\alpha}{\lambda} \left[ N_X \Delta_3 \langle \cos^4{\phi_X}\rangle-N_Y \Delta_4 \langle \cos^4{\phi_Y}\rangle\right]
\label{d1 d2}
\end{eqnarray}
of Eq.~(\ref{d1 d2}) can be computed from Eqs.(\ref{X equation},\ref{Y equation}), and we get the following relation,
 \begin{eqnarray}
\Delta_2 &=&\dfrac{N_h \alpha}{\lambda^2}\left[\Delta_1\left(g_1+\alpha \langle\cos{2\phi_h}\rangle g_0\right)+\Delta_2\left(g_1 \langle\cos{2\phi_h}\rangle+\alpha \langle\cos^2{2\phi_h}\rangle g_0\right)\right]
\label{g1 eq}
\end{eqnarray}
One rearranges Eq.~(\ref{g1 eq}) to find the ratio,
\begin{eqnarray}
\dfrac{\Delta_2}{\Delta_1}&=&\alpha N_h \dfrac{g_1+\alpha \langle \cos{2\phi_h}\rangle g_0}{\lambda^2-\alpha N_h \left( g_1 \langle \cos{2\phi_h}\rangle+\alpha \langle \cos^2{2\phi_h}\rangle g_0 \right)}\nn\\
&=& 2 \alpha \dfrac{g+\alpha \langle \cos{2\phi_h}\rangle }{\left(1-\alpha^2  \langle \cos^2{2\phi_h}\rangle \right)+D},
\label{ratio calculation}
\end{eqnarray}
where
 \begin{equation}
D=\sqrt{4 \alpha^2 \left(\langle \cos{2\phi_h}\rangle^2-\langle \cos^2{2\phi_h}\rangle\right)\left(1-g^2\right) +\left(1+2 \alpha \langle \cos{2\phi_h}\rangle g+\alpha^2 \langle \cos^2{2\phi_h}\rangle \right)^2},
\end{equation} 
and
\begin{equation}
g(\Phi_e)=\dfrac{g_1}{g_0}=\dfrac{N_X \langle \cos^4{\phi_X}\rangle- N_Y \langle \cos^4{\phi_Y}\rangle}{N_X \langle \cos^4{\phi_X}\rangle- N_Y \langle \cos^4{\phi_Y}\rangle}.
\end{equation}
  Even though nematic order couples $s$- and $d$-wave symmetry and brings angular dependence to the superconducting gap function in the primary $s$-wave state, one finds that $\frac{\Delta_2}{\Delta_1}=0$ when the numerator of Eq.~(\ref{ratio calculation}) vanishes,
\begin{equation}
\left(N_X \langle \cos^4{\phi_X}\rangle- N_Y \langle \cos^4{\phi_Y}\rangle\right) +\alpha \langle \cos{2\phi_h}\rangle \left(N_X \langle \cos^4{\phi_X}\rangle + N_Y \langle \cos^4{\phi_Y}\right)=0.
\end{equation}
For this case,  the gap function on the hole pocket becomes purely s-wave,  despite the presence of nematic order.

\end{widetext}
\end{document}